\documentclass[12pt,preprint]{aastex}
\usepackage{rotating,amsfonts,epsfig, amsmath,enumerate,color}
\begin{document}

\bibliographystyle{apj}
%\preprint{SPECTRUM.TEX v13}
   
\title{A Bayesian Approach to Classifying Supernovae With Color}
\author{
Natalia Connolly\altaffilmark{1},
Brian M. Connolly\altaffilmark{2}
}

\altaffiltext{1}{Physics Department, Hamilton College, Clinton, NY 13323, USA; nconnoll@hamilton.edu}
\altaffiltext{2}{Department of Physics and Astronomy, University of Pennsylvania, Philadelphia, PA 19104-6396, USA; brianco@sas.upenn.edu}

\email{}
\date{\today}% It is always \today, today,

%%%%%%%%%%%%%
\begin{abstract}
%%%%%%%%%%%%%
Upcoming large-scale ground- and space- based supernova surveys will face 
a challenge identifying supernova candidates largely without the use of spectroscopy.
Over the past several years, a number of supernova identification schemes have
been proposed that rely on photometric information only.  Some of these schemes use color-color
or color-magnitude diagrams; others simply fit supernova data to models.  Both
of these approaches suffer a number of drawbacks partially addressed in the so-called
Bayesian-based supernova classification techniques.  However, Bayesian techniques are 
also problematic in that they typically require that the supernova candidate
be one of a known set of supernova types.  This presents a number of problems, 
the most obvious of which is that there are bound to be objects that do not conform
to any presently known model in large supernova candidate samples.  
We propose a new photometric classification scheme that uses a Bayes factor
based on color in order to identify supernovae by type.
This method does not require knowledge of the complete set of possible 
astronomical objects that could mimic a supernova signal.
Further, as a Bayesian approach, it accounts for all
systematic and statistical uncertainties of the measurements 
in a single step.  To illustrate the use of the technique, we 
apply it to a simulated dataset for a possible future large-scale space-based
Joint Dark Energy Mission and demonstrate how it could be used to identify Type Ia supernovae.
The method's utility in pre-selecting
and ranking supernova candidates for possible spectroscopic follow-up -- \emph{i.e.}, 
its usage as a supernova trigger -- will be briefly discussed.  
\end{abstract}


%%%%%%%%%%%%%%%%%%%%%%%%%%%%%%%%%%%%%%%%%%%%%
\section{Introduction}
\label{section:intro}
%%%%%%%%%%%%%%%%%%%%%%%%%%%%%%%%%%%%%%%%%%%%%
In recent years, the question of photometric identification 
of supernova candidates has emerged as one of the crucial issues to be
resolved before the advent of large-space supernova cosmology 
experiments, both ground-based (\emph{e.g.}, the Large Synoptic Survey Telescope [LSST], 
the Dark Energy Survey [DES], the Panoramic Survey Telescope and Rapid Response System [Pan-STARRS]),
and space-based (\emph{e.g.}, the \emph{Joint Dark Energy Mission [JDEM]}).   
There are a number of reasons 
for this.    First, although there have been some interesting developments
in the possible uses of supernova other than Type Ia for 
cosmology~\citep{Baron:2004,hamuy,nugent,poznanski},
Type Ia supernovae (SNIa) remain the staple
of experimental cosmology.  
Second, SNe Ia are most reliably identified
using spectroscopy due to the presence of a characteristic SiII line at
6150 \AA\, in the supernova rest frame.  However, future large ground-based
surveys are expected to collect thousands of supernova candidates, making
a spectroscopic follow-up of each candidate all but unrealistic.  
The identification of supernova candidates (with possible spectroscopic follow-up
for a select sample) based on broadband photometry
remains the only feasible alternative.  

There have been a number of methods proposed to identify supernovae
using broadband photometry that can be divided into three broad categories.
One includes methods that rely on color-color or color-magnitude diagrams
~\citep{pozn02,riess2004,john,sull1}.  It is also possible to 
fit supernova data to models, and select the best fit (using, for example,
a $\chi^2$), which can be used to represent the supernova type~\citep{jha,salt,conley}.
Finally, the third category involves recently developed techniques 
based on a probabilistic (Bayesian) approach to the problem~\citep{bib:kuzn,pozn07}.
The method proposed in this work, although closer in spirit to the second category, 
has a number of advantages over both.  

The existing techniques, while adequate in many cases, have a number 
of serious shortcomings.  For example, supernova identification schemes based 
on color-color and color-magnitude diagrams
involve comparing the colors and/or magnitudes of a given 
supernova candidate with what is predicted by various 
supernova models.  This is an intuitive approach, allowing one
to visually judge the goodness of fit of the data 
to the models; however, it is difficult
to account for all statistical and systematic uncertainties in a single step.

A class of techniques that could be generally described as ``$\chi^2$-based'' 
simply find the best fit for a given supernova candidate's light curves
to a supernova model.  This is also an intuitive and often completely
reasonable approach, which nevertheless suffers the following
disadvantages:
%%%%%%%
\begin{enumerate}
\item \label{item:1} 
``This object is not 
a supernova of any kind'' is not a well-defined hypothesis in this
as in any other frequentist approach~\citep{bib:ed}.

\item 
Conversely, if the data happen to have large uncertainties, 
there is the possibility that a number of supernova models
will be good fits to the data.  There is no formalism to 
compute not only the probability that a given fit is good, but also
that it is bad.  In other words, what one is interested in is
the posterior probability, the probability that a
given hypothesis is true given the data.  Calculating this probability
requires that the probability that this hypothesis is false be
also known.     

\item 
Using a $\chi^2$-based technique only gives the information 
about the best fit for a given set of data to a model, while any 
information about worse fits is lost.  The best fit
will not necessarily reflect the true properties of the supernova.

\item 
In cases where one would like to use a tail probability 
for accepting or rejecting given supernova candidates (\emph{e.g.}, as 
SNe Ia), the probability of
falsely rejecting the null hypothesis (the so called Type I error rate) 
can be shown to be severely underestimated (see \cite{sel01} and references therein).
\end{enumerate}
%%%%%%%%

Bayesian classification schemes address many of the problems of the above-mentioned
methods.  However, existing Bayesian-based supernova
typing methods have a serious drawback: they require the knowledge of the complete
set of objects that a supernova candidate might conceivably be  
\citep{bib:kuzn,pozn07}.  That is, they
assume that a supernova candidate can only be one of a finite set of supernova types.  
%In one attempt to account for objects that do not conform to any known
%supernova models within the Bayesian approach,
%Type II-P supernovae were used as proxies for unmodeled phenomena
%that may mimic a  supernova signal~\citep{pozn07}.
%However, it is not clear whether, with better data and more accurate templates, 
%such objects would remain good proxies.  
%In general, requiring a finite set
%of supernova types is obviously rather limiting, 
%since, with the expanding scopes of supernova surveys, it is highly likely that 
%many more never-before measured non-SNe Ia types might be discovered, especially 
%at high redshifts.  
However, even with the current small high-redshift SN sample
(obtained almost exclusively with the Hubble Space Telescope) one occasionally finds 
supernova candidates with surprising new properties that do not 
seem to conform to any known models~\citep{bib:barb}.
Problems with  assuming a finite set of supernova-like objects are 
further addressed in Section~\ref{sec:danger}.

In our work, we introduce a likelihood ratio (a Bayes factor) 
that is capable of discriminating between SNe Ia and \emph{anything else} 
based on broadband photometric measurements.
The most important feature of this technique is that it is 
independent of the knowledge of the complete set of objects that a 
supernova candidate might conceivably be.  Another advantage is that,
as with all Bayesian-based techniques, this method allows one
to include all possible statistical and systematic uncertainties in a single
step.  Finally, the Bayes factor is formulated in terms of color and 
thus does not require that one make any assumptions about        
the absolute magnitudes of the supernova candidates in the 
broadband filters used.  Of course it is often desirable to include magnitudes
in the formalism; however, not only does it require making assumptions about 
the distribution of magnitudes for various known supernova types, but also
it places a hard upper limit on the intrinsic magnitudes of objects that have yet
to be observed.  But more importantly, ``anomalous'' (non-supernova) 
objects can be defined
in a far more mathematically elegant and computationally manageable way
using color alone.

The Bayes factor is defined as
\begin{eqnarray}
\mathcal{R}=
P(\text{Phot}|\text{Ia})/P(\text{Phot}|\text{non-Ia}).
\label{eqn:general}
\end{eqnarray}
where $P(\text{Phot}|\text{Ia})$ is the probability of obtaining the observed photometry (colors)  
from a SN Ia, and $P(\text{Phot}|\text{non-Ia})$ is the probability  of obtaining the 
data for any other object (which could be a non-SN Ia or any other object capable of 
mimicking an SN Ia signal).  Both probabilities take into account the {\it relative} 
distribution of light among the broadband filters used for the measurements.
In general, no specific set of models (or templates) for non-SNe Ia is required for the calculation 
of the denominator. 

On a more technical note, it is worthwhile to point out that Bayes factors are normally used for
deciding on the best of two hypotheses.  This allows one to easily set thresholds on the Bayes
factor in terms of the so-called Type I and Type II error rates\footnote{Type I error is
the probability of rejecting the null hypothesis when the null hypothesis is in fact correct;
it is thus a measure of the purity of the selection.  Type II error is the probability
that the null hypothesis will be accepted when the null hypothesis is in fact false; it is thus
a measure of the efficiency of the selection} in the same way as thresholds
are set on the likelihood ratio in~\cite{Wald:1945,Wald:1947}.  Also, although the main
focus of this work is to describe a method that can identify SNe Ia, the Bayes
factor can be easily cast in terms
of a posterior probability that a candidate is a Type $T$ supernova,
where $T$ could be Ibc, II-P, IIn, \emph{etc.}. 
\footnote{Consider some data $D$,
a hypothesis $\mathcal{H}_0$, and its
alternative $\mathcal{H}_1$.  The
Bayes factor can be defined as
\begin{eqnarray}
\mathcal{R}=\frac{P(D|\mathcal{H}_1)}{P(D|\mathcal{H}_0)}.
\end{eqnarray}
The posterior probability that the alternative hypothesis is true for the data can then be
written in terms of $\mathcal{R}$ provided
that one knows the priors for $\mathcal{H}_0$
and $\mathcal{H}_1$, denoted by $P(\mathcal{H}_0)$ and $P(\mathcal{H}_1)$, respectively:
\begin{eqnarray}
P(\mathcal{H}_1|D)=\frac{\mathcal{R}\frac{P(\mathcal{H}_1)}{P(\mathcal{H}_0)}}
{\mathcal{R}\frac{P(\mathcal{H}_1)}{P(\mathcal{H}_0)}+1}.
\end{eqnarray}
See~\cite{Berger:2001} for details.
Although there are historical reasons why the 
Bayes factor is formulated in this way, it is 
also convenient when setting thresholds for the error
rates because often the errors rates are at least
somewhat determined by the information contained in the priors on $\mathcal{H}_0$ and 
$\mathcal{H}_1$.}

This paper is organized as follows.  In Section~\ref{section:math} we 
derive an expression for $\mathcal{R}$ for a number of different cases.
We describe the performance of the method in  Section~\ref{section:performance}.  
Section~\ref{section:disc} presents a discussion of the results.

%%%%%%%%%%%%%%%%%%%%%%%%%%%%%%
\section{Derivation of the Bayes Factor}
\label{section:math}
%%%%%%%%%%%%%%%%%%%%%%%%%%%%%%
\subsection{Overview of the Calculation}
\label{section:calc_overview}

The Bayes factor, $\mathcal{R}$, introduced above, is defined as the probability 
of obtaining the photometric measurements assuming that the supernova candidate
is a Type Ia over the probability that it is anything else.  In practice, the probability 
that a candidate is an SN Ia is the probability that the 
colors are consistent with what is expected for an SN Ia using some prior knowledge about the behavior of Type Ia's.  
In our first formulation of the Bayes factor, if the candidate
is not in fact an SN Ia, then the distribution of light in the  broadband filters  used
can be arbitrary.  However, one could argue that much of the background for SNe Ia 
will be supernovae of other types whose behavior is relatively well-known.
However, the unprecedented
scale of the future supernova surveys makes it highly likely that many new types of
transient objects will be discovered.  Also, little is known about the \emph{rates}
of non-Type Ia supernovae, especially at very high redshifts, making it difficult to predict the\
 behavior of the background at those redshifts.

We begin with a general overview of the calculation of $\mathcal{R}$.  For simplicity,
we assume that there are only two broadband filters, and that there is a single measurement 
of the supernova candidate's flux in each.  Suppose that the flux is measured in 
photon counts, and that $M_1$ counts are measured in the first filter, and $M_2$, in the 
second.  Further suppose that there exists a model (a template) for the behavior
of SNe Ia in these filters, and that the model predicts that
some mean fraction of photons, $\bar{f}$, must end up in the first filter, 
and 1-$\bar{f}$, in the second.  The numerator of  $\mathcal{R}$,
$P(\text{Phot}|\text{Ia})$, is essentially the probability that 
the measurement is consistent with this model.  Assuming  
Poisson statistics for the photon distributions, it can be easily shown 
that $P(\text{Phot}|\text{Ia})$ takes the form of a 
standard binomial distribution:
\begin{eqnarray}
P(\text{Phot}|\text{Ia}) = \binom{M_1+M_2}{M_1} \bar{f}^{M_1}(1-\bar{f})^{M_2}.\notag
\end{eqnarray}  
For the calculation of the denominator of  $\mathcal{R}$, 
$P(\text{Phot}|\text{non-Ia})$, we do not make any {\it a priori} assumptions about 
the fraction of light that will end up in either filter.  The Bayesian framework
of the calculation allows one to circumvent this difficulty by marginalizing, or
integrating over, all possible fractions.  Mathematically,
\begin{eqnarray}
P(\text{Phot}|\text{non-Ia}) = \int_0^1 df \binom{M_1+M_2}{M_1} f^{M_1}(1-f)^{M_2}.\notag
\end{eqnarray}  

In reality, the calculation becomes rather more complicated.  To begin with, the measured
flux will most likely be better described using Gaussian, rather than Poisson, statistics.
We must also allow for the possibility of multiple measurements and more than 2 filters.
In the next section we will make the the calculation more explicit and account for all
of these factors.

%%%%%%%
\subsection{Mathematical Details}
\label{section:details}
%%%%%

Before we plunge into the full derivation of $\mathcal{R}$ for the case
of Gaussian statistics and multiple measurements and filters, 
we take a closer look at the simple case of a single measurement
of a supernova candidate in just two filters, assuming that the 
photon count fluctuations in the filters are Poisson.  Recall that we assume
that $M_1$ counts are measured by the first filter and $M_2$ by the second;
and that we have a model that predicts a certain fraction of 
photons, $\bar{f}$, for the first filter, and 
$(1-\bar{f})$, for the second. 

Following a similar derivation in \cite{jeffreys},
let us now introduce two variables, $f$ and $b$, such that  
the mean number of photons in the first filter is given by 
$fb$, and the mean number of photons in the second filter is
given by $(1-f)b$.   Variable $b$ ranges from  0 to $\infty$, and
can be thought of as the mean number
of photons that are counted in both filters for a given measurement.
Variable $f$ ranges from 0 to 1, and can be thought of 
as the probability that the photons will end up in the first
filter as opposed to the second.  An analogy would be  
collecting balls into two receptor bins with 
different volumes: in this case, $b$ would be the mean number 
of balls that will enter both bins, and $f$ is the 
relative ``acceptance'' of one bin.  The introduction of these
variables allows us to expand the Bayes factor, Eqn.\,\ref{eqn:general},
in terms of $f$ and $b$:
\begin{eqnarray}
\mathcal{R}
=\frac{\int_0^\infty db P(\text{Phot}|b,\text{Ia})P(b|\text{Ia})}
{\int_0^1df \int_0^\infty db P(\text{Phot}|f,b,\text{non-Ia})P(f,b|\text{non-Ia})}.
\label{eqn:general_expanded}
\end{eqnarray}

Here, the first term in the numerator, 
$P(\text{Phot}|b,\text{Ia})$, is the likelihood 
of obtaining the measurement given that the mean number of photons was 
measured to be $\bar{f}b$ in the first filter, and $(1-\bar{f})b$ in the second.  
Likewise, the first term in the denominator, $P(\text{Phot}|f,b,\text{non-Ia})$,
is the likelihood of obtaining the measurement given that the mean number of photons was 
measured to be $fb$ in the first filter, and $(1-f)b$ in the second.  
Note that the numerator is not a function of $f$ because $f$ is single valued
in the numerator, $f=\bar{f}$.  If the photon
distribution is governed by Poisson statistics, then:
\begin{eqnarray}
P(\text{Phot}|b,\text{Ia})
=\frac{(\bar{f}b)^{M_1}e^{-\bar{f}b}}{M_1!}
\frac{((1-\bar{f})b)^{M_2}e^{-(1-\bar{f})b}}{M_2!}.
\label{eqn:photo_like1}
\end{eqnarray}
and
\begin{eqnarray}
P(\text{Phot}|f,b,\text{non-Ia})
=\frac{(fb)^{M_1}e^{-fb}}{M_1!}
\frac{((1-f)b)^{M_2}e^{-(1-f)b}}{M_2!}.
\label{eqn:photo_like2}
\end{eqnarray}

The terms $P(b|\text{Ia})$ in the numerator and $P(f,b|\text{non-Ia})$ in 
the denominator of Eqn.\,\ref{eqn:general_expanded}, are prior probabilities
containing information regarding
the expected distribution of light in the two filters for an SN Ia 
and anything else, respectively.  Defining $b_{min}$ and $b_{max}$ as 
the minimum and maximum bounds for $b$ and assuming each value for $b$ in between
these bounds is equally probable, we have:
\begin{eqnarray}
P(b|\text{Ia})=\frac{1}{b_{max}-b_{min}}.
\end{eqnarray}
Note that the range of $b$ will always be assumed to be $b=[0,\infty]$,
although the upper and lower bounds will initially 
be set to $b_{max}$ and $b_{min}$, respectively.
\footnote{Here we have adopted Jaynes' methodology~\citep{Jaynes:2003}, 
expressing the prior probabilities in terms of variables representing the
bounds of those variables.  These bounds are 
inserted at the end of the calculations (integrations) with the goal
of avoiding handling variables whose limits are defined as $[0,\infty)$ -- 
\emph{i.e.}, to avoid priors whose probabilities approach 0.} 
However, if the candidate is not an SN Ia, we do not make 
any assumptions about what to expect, and so:
\begin{eqnarray}
P(f,b|\text{non-Ia})=\frac{1}{(b_{max}-b_{min})(f_{max}-f_{min})}=\frac{1}{b_{max}-b_{min}}.
\end{eqnarray}
as the upper ($f_{max}$) and lower ($f_{min}$) bounds of $f$ 
are 1 and 0, respectively.

Note that $P(b|\text{Ia})$ and $P(f,b|\text{non-Ia})$ 
are \emph{improper priors} (in other words, they assume probability density functions that
are flat and are integrated from zero to infinity).
This is not a major issue for our calculation because the priors happen to cancel.
However, in general the use of improper priors must be treated with caution~\citep{Berger:2001}.  It is therefore important to check that $\mathcal{R}$ does indeed 
behave properly; this will be
addressed further in Section\,\ref{section:results}.

With these priors, Eqn.~\ref{eqn:general_expanded} becomes:
\begin{eqnarray}
\mathcal{R}
=\frac{\binom{M_1+M_2}{M_1} \bar{f}^{M_1}(1-\bar{f})^{M_2} }{\frac{1}{M_1+M_2+1} }.
\label{eqn:intermediate1}
\end{eqnarray}

In the calculation of Eqn.\,\ref{eqn:intermediate1} $b$ is effectively 
unconstrained, leaving the supernova candidate's magnitude free to take on any value.
That is,  as we are only concerned with the relative fraction of photons in 
each filter (\emph{i.e.}, color), 
we need not make any assumptions about the behavior of $b$.

We would now like to derive an equation analogous to Eqn.~\ref{eqn:intermediate1},
but for the case of Gaussian statistics.  Let us suppose that 
instead of measuring $M_1$ photons in the first filter and $M_2$ in the second, we
now measure a flux $F_1$ in the first filter with an error $\sigma_1$,
and a flux $F_2$ in the second filter with an error $\sigma_2$.  
As before, we parametrize the mean (or ``true'') fluxes in the two filters
as $fb$ and $(1-f)b$, and expand $\mathcal{R}$ in terms of $f$ and $b$, leading
to Eqn.~\ref{eqn:general_expanded}.  
We then simply replace 
the Poisson distributions with Gaussian ones using the usual notation for a Gaussian distribution,
$G(x;\mu,\sigma)=\frac{1}{\sqrt{2\pi}\sigma}{\rm e}^{-\frac{(\mu-x)^2}{2\sigma^2}}$.
Equation\,\ref{eqn:general_expanded} becomes
%the likelihoods $P(\text{Phot}|b, {\rm Ia})$ and $P(\text{Phot},|f, b,\rm{non-Ia})$ 
%are now given by
%\begin{eqnarray}
%P(\text{Phot}|b,{\rm Ia})&\equiv&P(F_1,F_2,\sigma_1,\sigma_2|b,{\rm Ia}) \nonumber \\
%&=&G(F_1;\bar{f}b,\sigma_1)G(F_2;(1-\bar{f})b,\sigma_2)
%\end{eqnarray}
%and 
%\begin{eqnarray}
%P(\text{Phot}|f,b,{\rm non-Ia})&\equiv&P(F_1,F_2,\sigma_1,\sigma_2|f, b, {\rm non-Ia})\nonumber \\
%&=&G(F_1;fb,\sigma_1)G(F_2;(1-f)b,\sigma_2)
%\end{eqnarray}
%respectively.
%the Gaussian analog of Eqn.\, becomes:
\begin{eqnarray}
\mathcal{R}
&=& \frac{\int_0^\infty db G(F_1;\bar{f}b,\sigma_1)G(F_2;(1-\bar{f})b,\sigma_2)}
{\int_0^1df \int_0^\infty db G(F_1;fb,\sigma_1)G(F_2;(1-f)b,\sigma_2)}.
\label{eqn:gaussian}
\end{eqnarray}
%which then reduces to:
%\begin{eqnarray}
%\label{eqn:gaussian}
%\mathcal{R}&=&
%\frac
%{
%\frac{1}
%{4\sqrt{\pi} \sigma_1\sigma_2\left[\frac{\bar{f}^2}{2\sigma_1^2}+\frac{(1-\bar{f})^2}{2\sigma_2^2}\right]}
%\text{exp}\left[\frac{\left(\frac{F_1\bar{f}}{\sigma_1^2}+\frac{F_2(1-\bar{f})}{\sigma_2^2}\right)^2}
%{4\left(\frac{\bar{f}^2}{2\sigma_1^2}+\frac{(1-\bar{f})^2}{2\sigma_2^2}\right)}
%-\frac{F_1^2}{2\sigma_1^2}
%-\frac{F_2^2}{2\sigma_2^2}\right]
%}
%{
%\int_0^1df \frac{1}
%{4\sqrt{\pi} \sigma_1\sigma_2\left[\frac{f^2}{2\sigma_1^2}+\frac{(1-f)^2}{2\sigma_2^2}\right]}
%\text{exp}\left[\frac{\left(\frac{F_1f}{\sigma_1^2}+\frac{F_2(1-f)}{\sigma_2^2}\right)^2}
%{4\left(\frac{f^2}{2\sigma_1^2}+\frac{(1-f)^2}{2\sigma_2^2}\right)}
%-\frac{F_1^2}{2\sigma_1^2}
%-\frac{F_2^2}{2\sigma_2^2}\right]
%}\notag\\
%&&\left[\frac
%{
%1-\text{erf}\left(\frac{\frac{F_1\bar{f}}{\sigma_1^2}+\frac{F_2(1-\bar{f})}{\sigma_2^2}}
%{2\sqrt{\frac{\bar{f}^2}{2\sigma_1^2}+\frac{(1-\bar{f})^2}{2\sigma_2^2}}}\right)
%}
%{
%1-\text{erf}\left(\frac{\frac{F_1f}{\sigma_1^2}+\frac{F_2(1-f)}{\sigma_2^2}}
%{2\sqrt{\frac{f^2}{2\sigma_1^2}+\frac{(1-f)^2}{2\sigma_2^2}}}\right)}\right]
%\end{eqnarray}
The integration over $b$ in Eqn.~\ref{eqn:gaussian} can be 
reduced further leading
to the appearance of the Gauss error function.  However, the integration over $f$ 
in the denominator can only be done numerically.

We now make Eqn.~\ref{eqn:gaussian} even more realistic by considering multiple
measurements (say, $N$) and an arbitrary number of filters (say, $M$).
Using the formalism we have developed above, we will assume that for 
the $j^{th}$ measurement the fraction of light in the $k^{th}$ filter
is $f^{\prime\,k} _j$, and the total light distributed between all the filters
and all the measurements is given by $b$.   Therefore, the hypothesized flux in the $k^{th}$ filter 
and $j^{th}$ measurement is given by  $f^{\prime\,k}_j\, b$.  Again, if the supernova
is assumed to be a Type Ia, then we have a model that describes the fraction 
of light in each of the filters for each of the measurements must be.  The model
must take into account the many possible observational parameters that 
characterize an SN Ia.  For example, it is known that  SNe Ia
have a variety of possible ``stretch'' values, which parametrize the width 
of their light curves~\citep{bib:perl}.  Following the approach used in~\cite{bib:kuzn}, 
we represent the Type Ia supernova
parameters by $\vec{\theta}$, defined as
\begin{eqnarray}
\vec{\theta} \equiv (s,A_v,R_v,t_{diff},z).  
\end{eqnarray}
where $s$ is the stretch parameter;  $A_v$ and $R_v$ parametrize the effect of interstellar dust 
extinction using the Cardelli-Clayton-Mathis (CCM) parametrization~\citep{bib:ccm}; 
$t_{diff}$ accounts for the difference in the time of maximum of 
the model and the data; and $z$ is the redshift.  In other words, for Type Ia's
$f^{\prime\,k} _j$ will become $f^{\prime\,k} _j(\vec{\theta})$.
The exact assumptions about the distribution of these parameters will be discussed below 
when the prior probabilities for all of the $\vec{\theta}$ parameters will be stated explicitly.

Since, in general, the exact values of each of these parameters for a given 
candidate are unknown, they must be marginalized.  
If the $i^{th}$ measurement in filter $k$ of the flux is $F_i^k$ with error $\sigma^k_i$,
the multi-measurement, multi-filter analog of Eqn.~\ref{eqn:general_expanded} is:
\begin{eqnarray}
\mathcal{R}&=&
\frac
{\sum_{\vec{\theta}} \int_{0}^{\infty} db {P(\{F_i^k\},\{\sigma_i^k\}|{\vec{\theta}} ,b,{\rm Ia})P({\vec{\theta}} ,b|\text{Ia})}}
{ \int_{0}^{1} d{\bf f}^\prime\int_{0}^{\infty} db {P(\{F_i^k\},\{\sigma_i^k\}|{\bf f}^\prime ,b, {\rm non-Ia})P({\bf f}^\prime ,b|\text{non-Ia})}}.
\label{eqn:terms}
\end{eqnarray}
where ${\bf f}^{\prime}=\{f_{j}^{\prime\,k}\}$, and $\int_{0}^{1} d{\bf f}^{\prime}$
indicates an integration over the multi-dimensional parameter
space where $\sum_{k=1}^M\sum_{j=1}^N f_{j}^{\prime\,k}=1$.
The denominator is not parametrized by ${\vec{\theta}}$ as it is 
not known what parameters are relevant for what we define as 
``anything other than SNe Ia''.  Therefore, every possible distribution of light in
the filters is given an equal chance.

We now address each term in Eqn.~\ref{eqn:terms} in turn.  
$P(\{F_i^k\},\{\sigma_i^k\}|{\vec{\theta}} ,b, {\rm Ia})$ \\
 and $P(\{F_i^k\},\{\sigma_i^k\}|{\bf f}^\prime ,b, {\rm non-Ia})$
are the likelihoods of 
obtaining a set of fluxes, $\{F_i^k\}$, with uncertainties $\{\sigma_i^k\}$,
for a number of measurements and filters, given that the mean number 
of photons are measured to be $\{f_j^{\prime\,k}({\vec{\theta}}) \, b\}$
and $\{f_j^{\prime\,k} \, b\}$, respectively.
Assuming each measurement and every filter are independent,
\footnote{Note that 
as long as the overlap between the filters is not 100\%,
then, without assuming anything about the underlying spectrum,
any relative fraction of light is allowed between the two filters.
}
\begin{eqnarray}
P(\{F_i^k\},\{\sigma_i^k\}|{\vec{\theta}}, b, {\rm Ia})=
\prod_{k=1}^M\, \prod_{i=1}^N 
G(F_i^k;f_{j}^{\prime\,k}({\vec{\theta}}) b,\sigma_{i}^{k}).
\end{eqnarray}
and
\begin{eqnarray}
P(\{F_i^k\},\{\sigma_i^k\}|\{f_j^{\prime\,k}\},b, {\rm non-Ia})=
\prod_{k=1}^M\, \prod_{i=1}^N 
G(F_i^k;f_{j}^{\prime\,k} b,\sigma_{i}^{k}).
\end{eqnarray}
Note that, in general, the measured flux ($F_i^k$) and the hypothesized 
flux ($f^{\prime\,k} _j b$) have different subscripts (which indicate the 
measurement number).  This is done to emphasize the fact that it is unknown where the time of maximum of our measured 
light curve is relative to that of the model.  This uncertainty is taken into 
account in one of the $\vec{\theta}$ parameters, $t_{diff}$.

The terms $P({\vec{\theta}},b|\text{Ia})$ and $P({\bf f}^\prime,b|\text{non-Ia})$  in Eqn.~\ref{eqn:terms}
are prior probabilities.  In particular, $P({\vec{\theta}},b|\text{Ia})$ contains
the prior knowledge about the parameters $\vec{\theta}$ that describe an SN Ia.
For $P({\bf f}^\prime,b|\text{non-Ia})$, ${\bf f}^\prime$ is not constrained.
Likewise, parameter $b$ is not constrained 
in any way for either a Type Ia or a non-Ia prior.  It is therefore
marginalized.  Integrating over $b$ means integrating
over the total light in all the filters and all the measurements for 
a given candidate -- \emph{i.e.}, integrating over the
observed magnitude for this measurement.  Furthermore, in marginalizing
$b$, there is an implicit assumption about the prior distribution of $b$ -- namely,
that it is flat.  This assumption allows us to formulate the 
probabilities purely in terms of color.
Allowing the total light to vary measurement-by-measurement with a flat prior 
is arguably the lightest possible assumption one can make 
regarding the magnitude.  
%That is a multinomial distribution
%is derived from integrating many Poisson distributions whose
%means are fractions of total mean
%(see Appendix\,[]).  

Explicitly, the priors $P(\vec{\theta},b|\text{Ia})$ and 
$P({\bf f^\prime},b|\text{non-Ia})$ become:
\begin{eqnarray}
P({\bf f},b|\text{non-Ia})=\frac{1}{b_{max}-b_{min}}\delta\left(1-\prod_{j=1}^N  \sum_{k=1}^M f_{j}^{\prime\,k}\right),
\end{eqnarray}
since the only constraint here is that all the light fractions in different filters add up to one; and 
\begin{eqnarray}
P(\vec{\theta},b|\text{Ia})=\xi(\vec{\theta}) \frac{1}{b_{max}-b_{min}}
\end{eqnarray}
where  
$\xi(\vec{\theta})$ is a the prior probability of $\vec{\theta}$.

The priors on $\vec{\theta}$ are defined similarly to 
those in~\cite{bib:kuzn} and are briefly summarized below.  
The stretch parameter $s$ follows a Gaussian distribution 
with a mean of $\bar{s}=0.97$ and a width of $\delta s =0.09$ 
(these values are extracted from~\cite{bib:sull}).  The CCM parameters
 $A_v$ and $R_v$ can assume two sets of values with equal probabilities:  
$(A_v,R_v)=(0.0,0.0)$ (no extinction) and $(0.2,2.1)$ (moderate extinction).
The prior probability for each choice of $A_v$ and $R_v$ is therefore $N_{dust}=1/2$.
The parameter accounting for the difference between the time of maximum of the data
and the model, $t_{diff}$, has a flat prior.  The measured light curve is shifted
relative to a template in one day increments 1000 times and each 
shift is assigned an equal probability $N_{t_{diff}}=1/1000$.
means that a flat prior is assigned to $t_{diff}$.
Finally, the redshift parameter $z$ is assumed to be known from 
the supernova candidate's host galaxy, $z_{gal}$, with
an associated uncertainty of $\sigma_{gal}$.
We consider the range of redshifts from 0 to 1.7, and 
assume two representative possibilities for  $\sigma_{gal}$,
0.005 (which might be obtained through a spectroscopic analysis of the
host galaxy's spectrum) and 0.1 (obtained through a photometric analysis).  Therefore,
\begin{eqnarray}
\xi(\vec{\theta}) = 
%\frac{1}{\sqrt{2\pi}\delta s}e^{-\frac{(s-\bar{s})}{2\delta s ^2}}\frac{1}{N_{dust}}\frac{1}{N_{t_{diff}}}
%\frac{1}{\sqrt{2\pi}\sigma_{gal}}e^{-\frac{(z-z_{gal})^2}{\sigma_{gal}^2}}
G(s;\bar{s},\delta s)\frac{1}{N_{dust}}\frac{1}{N_{t_{diff}}}
G(z_{gal};z,\sigma_{gal})
\label{eqn:thetas}
\end{eqnarray}

Putting everything together, we obtain the full Bayes factor:
\begin{small}
\begin{eqnarray}
\mathcal{R}&=&\label{eqn:exact_R} \frac
{\sum_{\vec{\theta}} \xi(\vec{\theta}) \int_{0}^{\infty} db \prod_{i=1}^N \prod_{k=1}^M  \, 
G(F_{i}^k;f_{i}^k(\vec{\theta})b,\sigma_{i}^k) }
{\int_{0}^{\infty} db \prod_{i=1}^N \prod_{k=1}^M\, \int_0^1 df_{j}^{\prime\,k} \,  G(F_{i}^k;f_{i}^kb,\sigma_{i}^k)\delta\left(1-\sum_{k=1}^M f_{j}^{\prime\,k}\right)}
\label{eqn:R_final}
\end{eqnarray}
\end{small}
where $\sum_{{\vec{\theta}}}$ represents the sums and integrations over the parameters
in ${\vec{\theta}}$ (depending on whether they are discrete or continuous).

Now the calculation of Eqn.~\ref{eqn:exact_R} 
requires performing $N\times M$ integrations over $f_{j}^{\prime\,k}$
in the denominator.  For a large number of filters (say, $\geq$ 8)
and many measurements, it is nearly impossible to do this calculation 
in a reasonable amount of time with the required precision for $f_{j}^{\prime\,k}$
without the use of techniques such as Markov Chain Monte Carlo integration methods.  
However, the number of integrations can be reduced to $M$ (the number of filters) 
if we allow for measurement-to-measurement variations in $b$.  That is, if we allow
\begin{eqnarray}
\label{eqn:b}
b\rightarrow b_i,
\end{eqnarray}
$b_i$ can be brought inside the product over measurements.  
This assumption is the equivalent of 
removing the knowledge that the colors between the 
measurements are known (or in the Poisson case, this is equivalent to 
having a separate multinomial for each measurement).  
This effectively
releases the constraint on colors between measurements.
Technically, the effect of this assumption should be to sweep more candidates which 
look less like SNe Ia into the SN Ia hypothesis, so 
if one's sample contains ``anomalous'' candidates that 
very closely mimic SNe Ia one might reconsider this
conjecture.

With Eqn.~\ref{eqn:b}, Eqn.\,\ref{eqn:R_final} becomes 
\begin{small}
\begin{eqnarray}
\mathcal{R}&=&\frac
{\sum_{\vec{\theta}} \xi(\vec{\theta})  \prod_{i=1}^N \int_{0}^{\infty} db_i \prod_{k=1}^M  \, 
G(F_i^k;f_{i}^{\prime\,k}(\vec{\theta}),\sigma_{i}^{k})}
{ \prod_{i=1}^N \int_{0}^{\infty} db_i \prod_{k=1}^M\, \int_0^1 df_{j}^{\prime\,k} \,  
G(F_i^k;f_{i}^{\prime\,k},\sigma_{i}^{k})\delta\left(1-\sum_{k=1}^M f_{j}^{\prime\,k}\right)}.
\label{eqn:R_final2}
\end{eqnarray} 
\end{small}

%Note only $M$ integrations over $f_{j}^{\prime\,k}$ for each of the measurements
%will be required.  

%\subsection{Including Magnitude}
%\begin{eqnarray}
%\mathcal{R}_{mag}=\frac{\sum_{\vec{\theta}}P(\text{Photo}|\vec{\theta},\text{Ia})P(\vec{\theta}|\text{Ia})}
%{\int_0^1\int_0^\infty df_j^{\prime\,k} db_i P(\text{Photo}|f_j^{\prime\,k},b_i,T,\text{non-Ia})
%P(f_j^{\prime\,k},b_i,T|\text{non-Ia})+
%\sum_T \sum_{\vec{\theta}} P(\text{Photo}|\vec{\theta},T,\text{non-Ia})
%P(\vec{\theta},T|\text{non-Ia})}
%\end{eqnarray}

%%%%%%%%%%%%%%%%%%%%%%%%
\section{Performance Studies}
\label{section:performance}
%%%%%%%%%%%%%%%%%%%%%%%%
\subsection{The Simulated Dataset}
\label{section:data}
%%%%%%%%%%%%%%%
In order to check the performance of the method proposed above, we simulate a dataset closely
mimicking one that could be obtained by a possible 
JDEM space-based mission.      
The mission is based on a 2-m class telescope, and is capable of taking multi-band photometric data in
the wavelength range from 0.3 to 1.7 $\mu$m.  Photometric data are assumed to be taken every 
4 days in the observer frame, with an exposure time of 1200 seconds.   
Note that this study is not meant to test the performance of any particular JDEM mission; 
we simply test the performance of the method assuming a fairly generic,
plausible JDEM.

We create a number of supernova candidates
of a given type using spectral templates from~\cite{bib:hsiao} for Type Ia's, 
and from P. E. Nugent for non-Ia's.\footnote{See http://supernova.lbl.gov/$\sim$nugent/nugent\_templates.html.}
The date of explosion for a given candidate is chosen randomly within the confines of a 
3 year mission timeline, and supernova candidate properties 
are generated according to the probability distribution functions
in $\xi(\vec{\theta})$ (see Eqn.~\ref{eqn:thetas}).
The supernova light curves are then realized using a simple 
aperture exposure time calculator.  We generate supernova candidates
of types Ia, Ibc, II-P, and IIn.    
We assume that the intrinsic rest-frame $B$-band magnitudes follow a Gaussian distribution,
with the mean and standard deviations obtained from~\cite{bib:rich}.
In particular, the mean and
standard deviation are taken to be $-19.05\pm 0.30$ mags. for Type Ia's;
$-17.27\pm 1.30$ mags. for Type Ibc's; $-19.05\pm 0.92$ mags. for Type IIn's and
$-16.64\pm 1.12$ mags. for Type II-P's.
We also generate ``anomalous'' objects by creating fake unimodal light curves
(light curves that rise and fall with a single maximum, but are otherwise
random).  These light curves are assigned 1\% errors in each broadband filter considered.

\begin{figure}[tbh]
\begin{center}
\epsfxsize=3.0in\epsffile{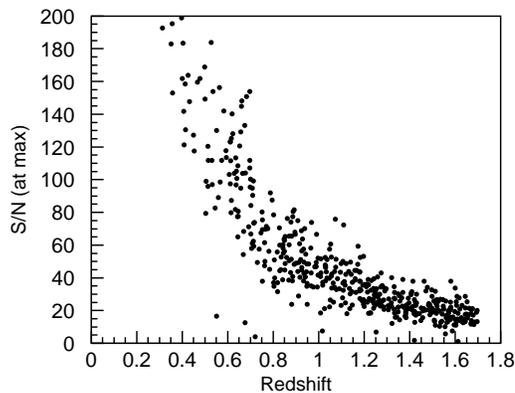} 
\end{center}
\caption {
Distribution of the signal-to-noise ratios at maximum light
as a function of redshift for the simulated SNe Ia, in the filter
that most closely matches the rest-frame $B$-band.
}
\label{fig:sn}
\end{figure}

To get a better feel for the simulated dataset, Fig.~\ref{fig:sn} shows the signal-to-noise ratios
at maximum light as a function of redshift for the simulated SNe Ia in the filter 
that most closely matches the rest-frame $B$-band.

Once the light curves are simulated, we select a subset of them 
in a limited number of broadband filters.
The chosen filters must include those that most closely match 
the rest-frame $B$- and $V$- bands for a given candidate. 
The reason why we place a particular emphasis on these bands
is because they correspond to a wavelength range where SNe Ia are particularly well modeled.
To limit the computational time, we only consider 50
consecutive photometric measurements
(the  actual available number of measurements varies evenly from 0 to over 350).
We also require that there be at least one measurement with a signal-to-noise 
ratio $>$ 5 in the filter most closely 
corresponding to the rest-frame $B$-band for a given candidate.  
Any space-based dark energy mission that extends to at least
a year and uses SNe Ia as a dark energy probe will satisfy these requirements
(in fact, every JDEM mission currently on the market does).

%This requirement 
%in fact makes little
%difference because we are using high-quality light curves with a sufficient number 
%of measurements to guarantee that this condition is nearly always satisfied.

%%%%%%%%
\subsection{Results}
\label{section:results}
\begin{figure}[tbh]
\begin{center}
\epsfxsize=3.0in\epsffile{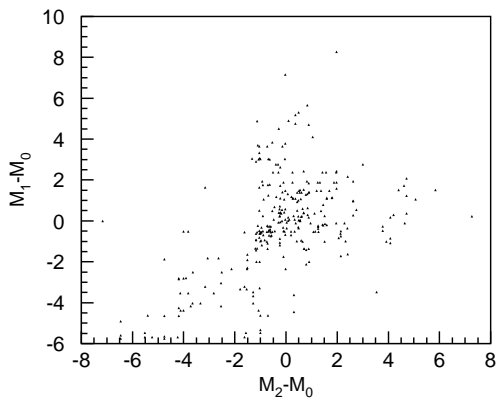} 
\end{center}
\caption {
The distribution of colors for the fake unimodal data.  $M_i$ is the magnitude in 
filter $i$ (filter 0 covers the range of 0.32-0.47 $\mu$m; filter 1, 0.41-0.56 $\mu$m; and filter 2,
0.49-0.68 $\mu$m).
}
\label{fig:fake_colors}
\end{figure}

\begin{figure}[tbh]
\begin{center}
\epsfxsize=3.0in\epsffile{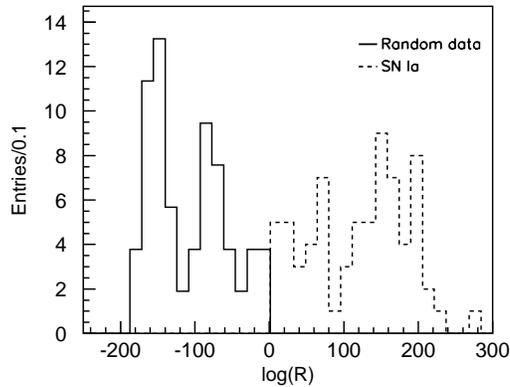} 
\end{center}
\caption {
Distributions of ${\rm log}\mathcal{R}$  for random unimodal data (solid line) 
and SN Ia's (filled histogram).   
The two histograms have been normalized to the same area for an easier shape comparison.
}
\label{fig:fakes}
\end{figure}

%%%%%%%%
A number of tests are used to check the performance of the method.
First, we calculate the Bayes factor, $\mathcal{R}$  (Eqn.~\ref{eqn:R_final2}), for a sample of 
simulated SNe Ia and a sample containing unimodal ``fake'' light curves.  
The unimodal light curves for a given ``object'' peak at the same time
in all the filter bands.  To give a sense
of their color distribution, Fig.~\ref{fig:fake_colors} shows the fake objects'
colors for the 3 lowest wavelength filters bands (the first filter covers the 
range of 0.32-0.47 $\mu$m; the second, 0.41-0.56 $\mu$m; and the third,
0.49-0.68 $\mu$m).
This test allows us to test the discrimination between SNe Ia and 
objects that are not supernovae of any kind.  Figure~\ref{fig:fakes} shows
that ${\rm log}\mathcal{R}$ is predominantly positive for SNe Ia and negative for the random,
unimodal data, meaning that
$\mathcal{R}<1$.  This means that the method behaves
as expected, discriminating between random data and true 
supernovae 100\% of the time.  

\begin{figure}[tbh]
\begin{center}
$\begin{array}{c@{\hspace{0.0in}}c}
\epsfxsize=3.0in\epsffile{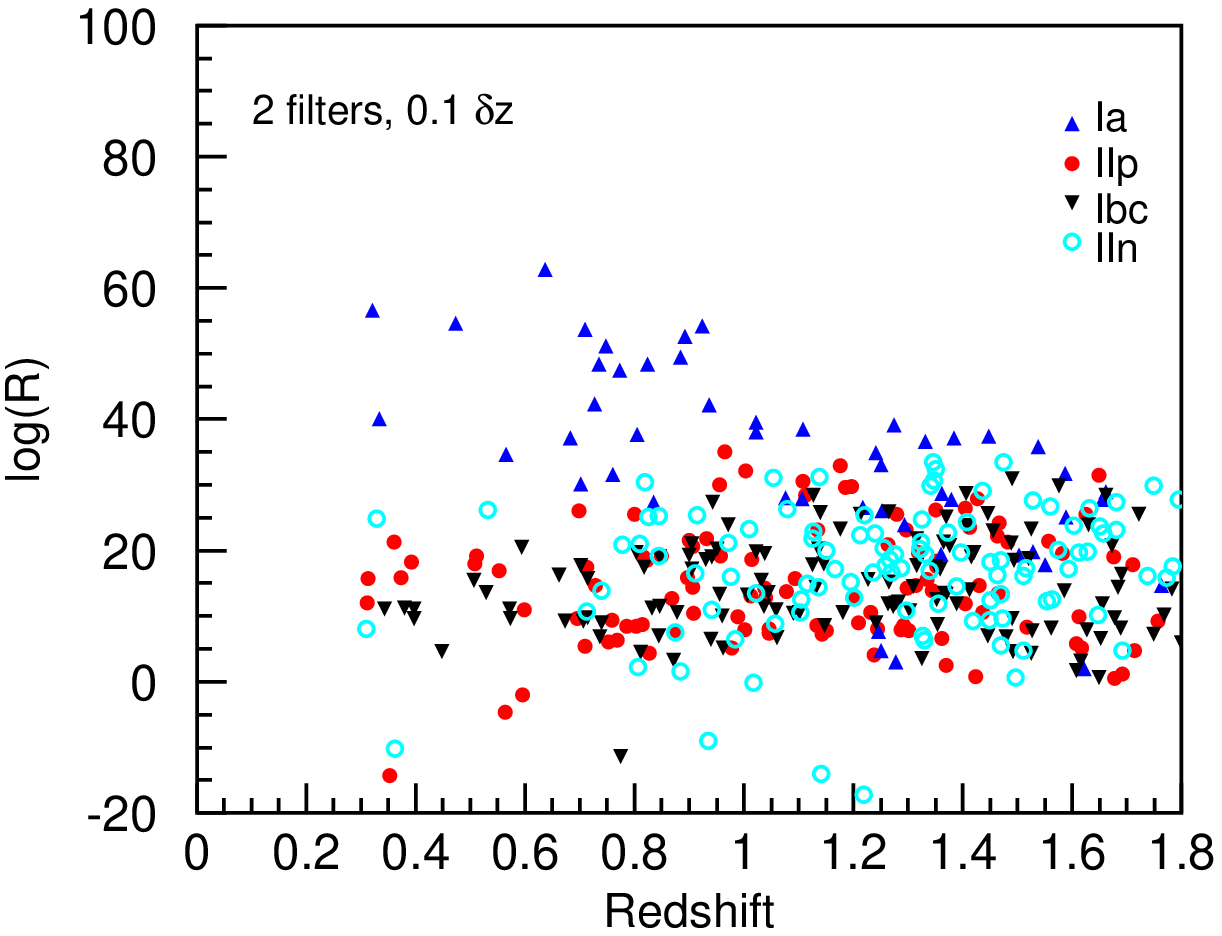} & 
\epsfxsize=3.0in\epsffile{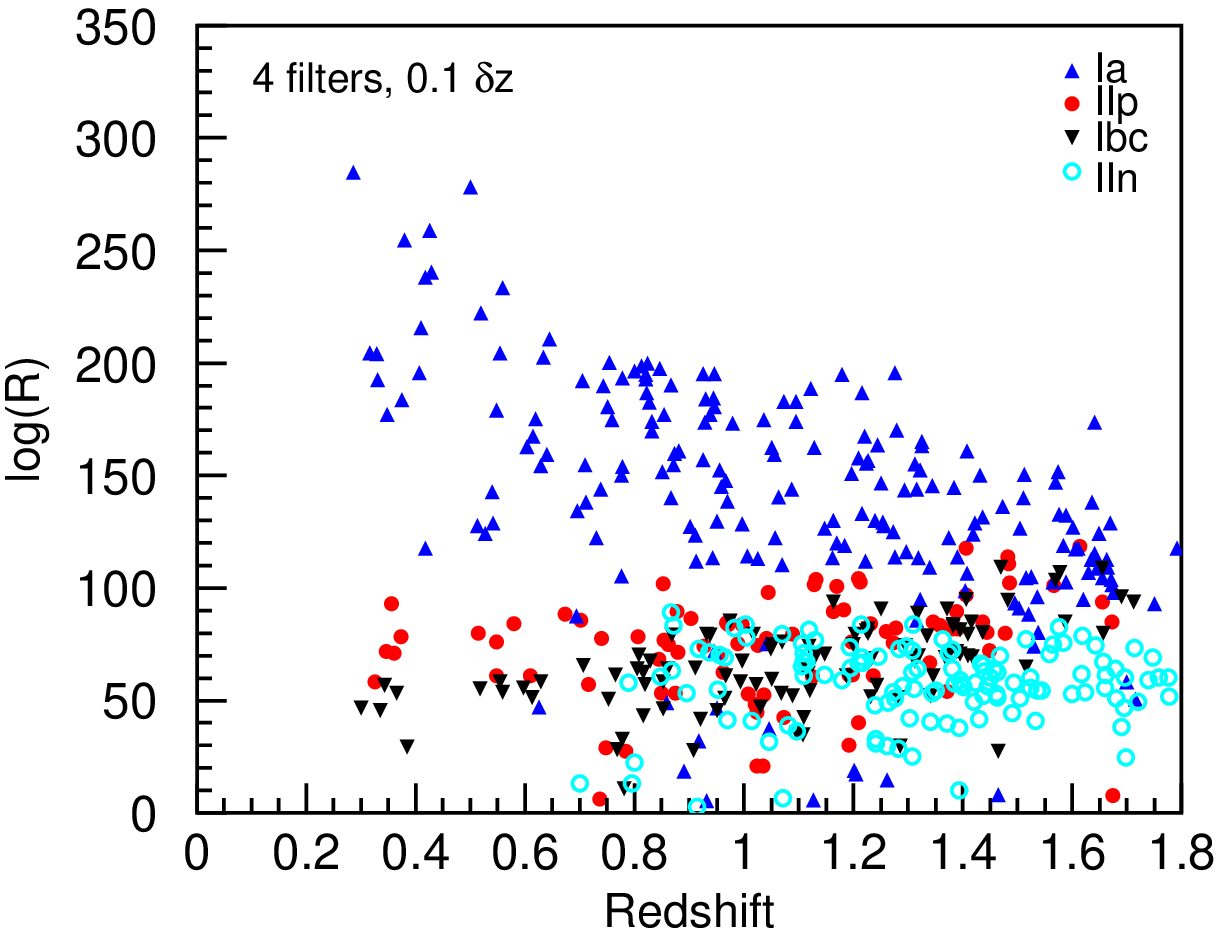} \\
\epsfxsize=3.0in\epsffile{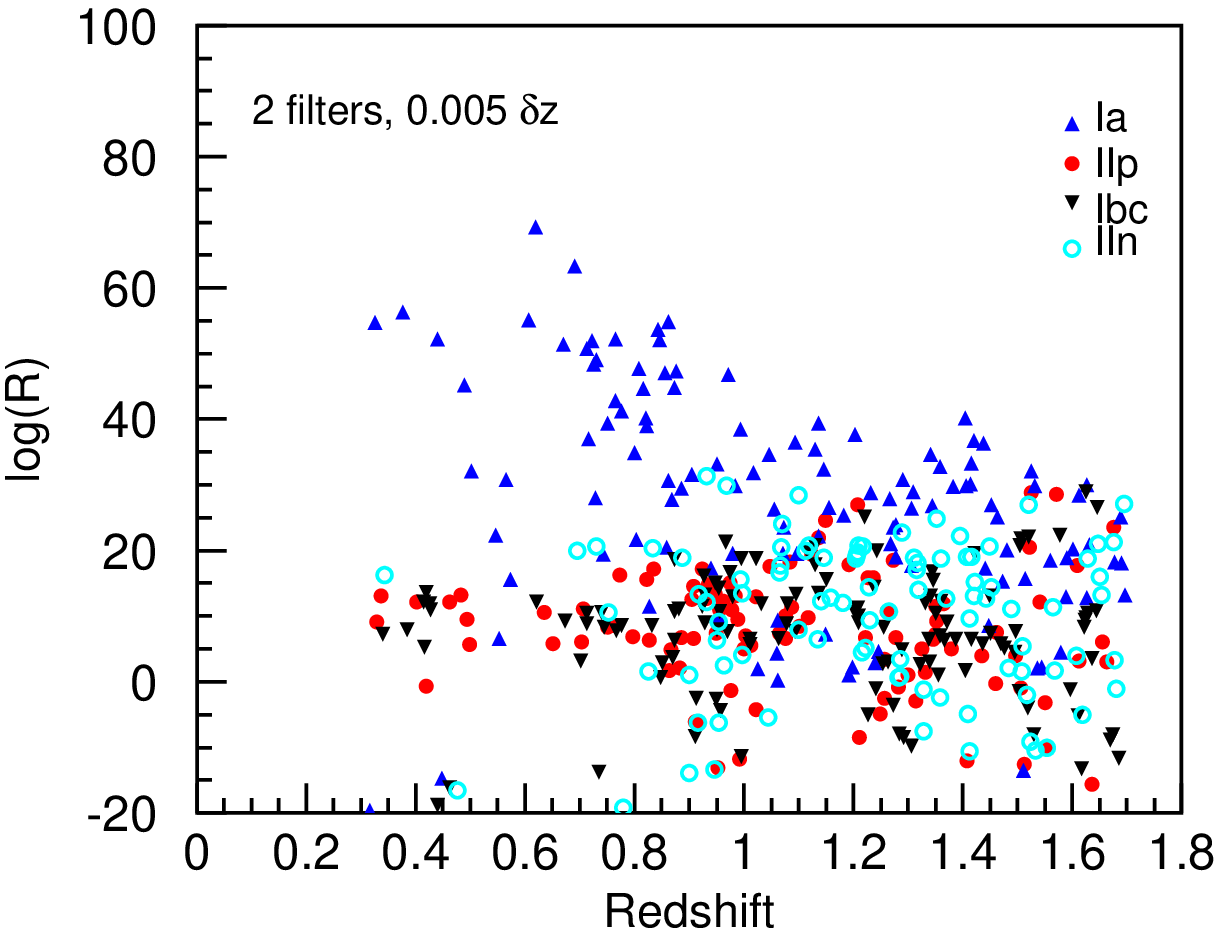} &
\epsfxsize=3.0in\epsffile{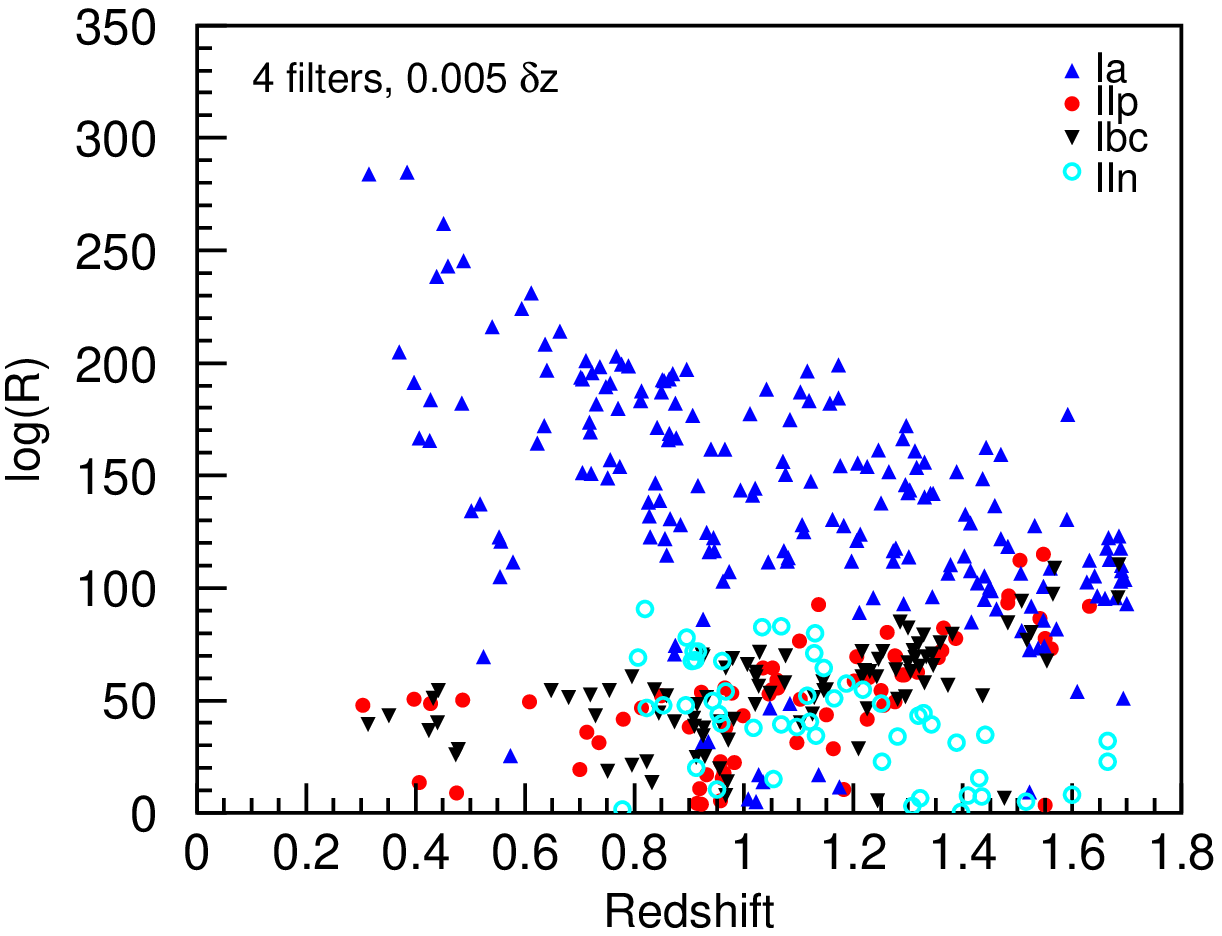} \\
\end{array}$
\end{center}
\caption {
Distributions of ${\rm log}\mathcal{R}$ vs. redshift for Type Ia's (upward turned triangles),
Type Ibc's (downward turned triangles),
Type II-P's (filled circles) and Type IIn's (open circles), with a 0.1 error
on the candidate redshifts (top row) and a 0.005 candidate
redshifts (bottom row), for 2 filters (left plots) and 4 filters (right plots).
The unimodal random data are not over-plotted on these figures because 
they all have large negative values $log(\mathcal{R})$'s 
that dwarf the $y$-axis scales.
}
\label{fig:IaandII-P}
\end{figure}

We also compute the Bayes factor for a sample
of simulated Type Ibc's, Type II-P's, and Type IIn's.   
%Figures~\ref{fig:dataandtempIa},
%\ref{fig:dataandtempIbc}, and\ref{fig:dataandtempII-P} show the fractions of light
%in 4 filters  for three typical simulated Type Ia, Ibc, and II-P supernova candidates, respectively; 
%as well as those for their respective models.  
%%%
%\begin{figure}[tbh]
%\begin{center}
%\epsfxsize=4.5in\epsffile{f3.ps}
%\end{center}
%\caption {
%The fraction of light for a  typical
%simulated SN Ia data (open circles with error bars) and for the corresponding SN Ia  
%template (solid line) in 4 broadband filters.  
%}
%\label{fig:dataandtempIa}
%\end{figure}
%%%
%\begin{figure}[tbh]
%\begin{center}
%\epsfxsize=4.5in\epsffile{f4.ps}
%\end{center}
%\caption {
%The fraction of light for a  typical
%simulated SN Ibc data (open circles with error bars) and for the corresponding  SN Ibc template
%(solid line) in 4 broadband filters.  
%}
%\label{fig:dataandtempIbc}
%\end{figure}
%%%
%\begin{figure}[tbh]
%\begin{center}
%\epsfxsize=4.5in\epsffile{f5.ps}
%\end{center}
%\caption {
%The fraction of light for a  typical
%simulated SN II-P data (open circles with error bars) and for the corresponding SN II-P 
%template (solid line) in 4 broadband filters.  
%}
%\label{fig:dataandtempII-P}
%\end{figure}
%%%
Note that we do not have to include any information about the expected
light curves for Type Ibc's, Type IIn's, or Type II-P's to compute  $\mathcal{R}$.  
Figure~\ref{fig:IaandII-P} 
shows the comparison of ${\rm log}\mathcal{R}$ for the 
case of measurements in 2 filters (left column) and 4 filters (right column).
We consider the case of a 0.1 error on the redshifts (top row) and a 0.005 error
on the redshift (bottom row).
For this comparison, we assume that the errors on the supernova fluxes are 
realistic (that is what a JDEM mission described above would
be expected to obtain).  
Several interesting conclusions can be drawn from Fig.~\ref{fig:IaandII-P}.
First, it is apparent that the method does discriminate between SNe Ia and 
the other types, although log$\mathcal{R}$ tend to be larger than 0 (so that
$\mathcal{R} > 1$) because SNe Ia are far more similar 
to other supernovae than they are to anything else. 
It should be noted that the values of $\mathcal{R}$ tends
to be quite large.  This is due to our use of a large number of measurements 
($\sim$ 50) in a number of filters, which ensures that a candidate is either
very much identified as an SN Ia-like candidate or not.
Second, as expected, the discrimination between SN Ia's and non-Ia's
increases with more information (4 filters vs. 2 filters) and/or with
better prior knowledge (\emph{i.e.}, smaller errors on the measured redshift).
Third, the discrimination is somewhat worse at high redshifts ($>$ $\sim$1), as we move
into a domain of less precise data and less certain models; but 
it is still good enough for $\mathcal{R}$ to be used as a first-pass SN Ia classifier.
Additionally, at least some plausible JDEMs ``sculpt'' their expected 
SNe Ia distribution so that it peaks at $z$ $\sim$ 0.7~\citep{bib:ald}.

\begin{figure}[tbh]
\begin{center}
\epsfxsize=3.0in\epsffile{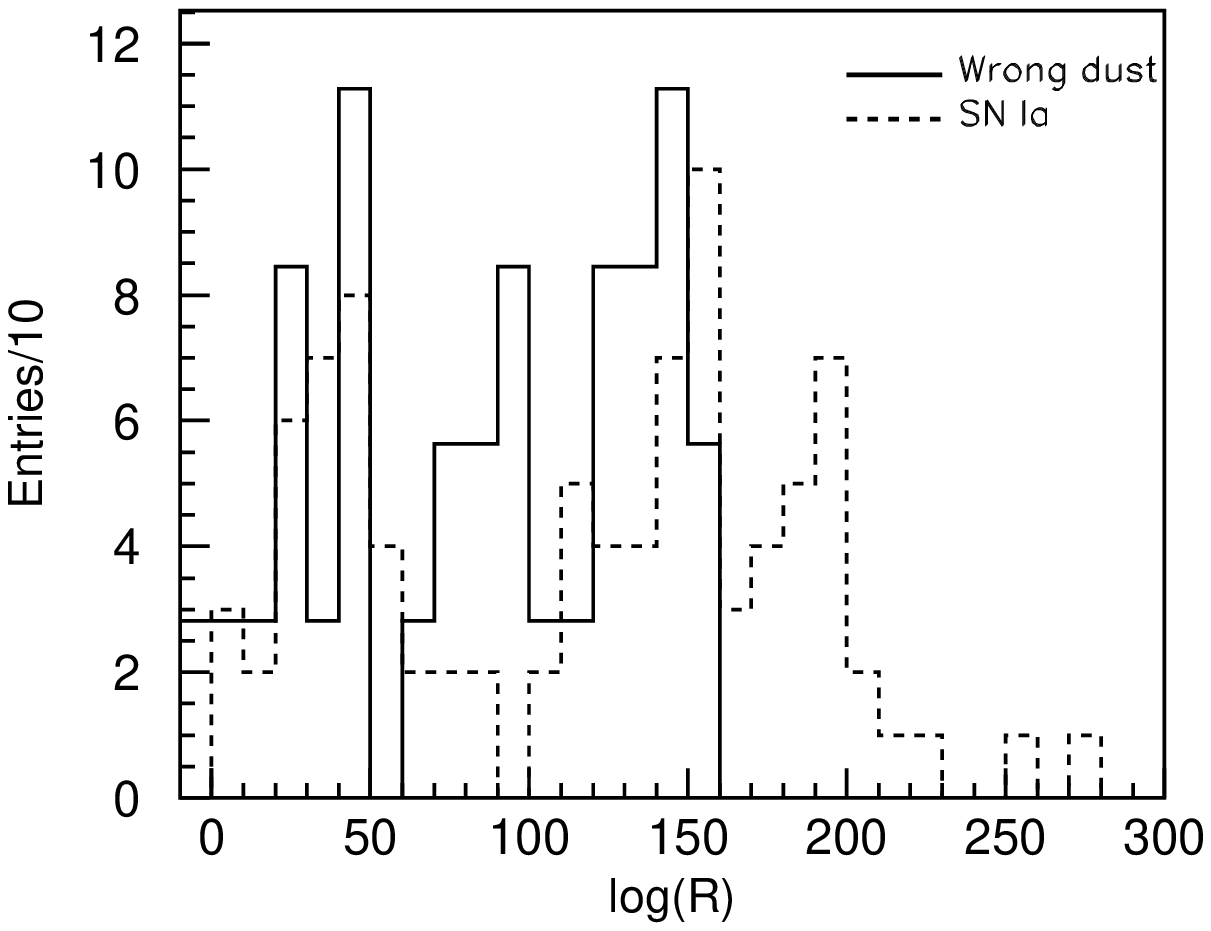}
\end{center}
\caption {The distributions of $\mathcal{R}$ for the 
case of the extinction mis-match between the data and the templates (solid line),
and the case of the matching extinction assumptions for the data and the templates
(dashed line).  The mis-matching case clearly results in a worse discrimination, 
making SNe Ia look more like ``anomalous'' objects; this is because the extinction
in the data is not accounted for in the set of templates used to define an SN Ia.
}
\label{fig:wrongDust}
\end{figure}

One might ask how  $\mathcal{R}$  would be affected if 
the templates used had incorrect colors for the SN Ia hypothesis.
In order to answer this question, we generated SN Ia candidates with $(A_v,R_v)=(0.2,2.1)$
but use only the no-extinction templates when calculating $\mathcal{R}$.  
Figure\,\ref{fig:wrongDust} shows the distributions of $\mathcal{R}$
for the cases where the extinction parameters in the data and 
templates are matched and mis-matched.
As expected,  $\mathcal{R}$ decreases when the extinction in the templates
does not match that in the data; that is, the SNe Ia look more
similar to non-SNe Ia.

%Regardless, there are a number of reasons why Type II-P's can be mistaken for Type Ia's.  
%First, most of the mis-identified Type II-P 
%candidates are those whose intrinsic magnitude has been dramatically fluctuated 
%down so as to make them appear almost as bright as Type Ia's.  One could argue that it is \emph{known} that Type II-P's 
%tend to be much less bright than Type Ia's.  However, if we are to take the errors on the intrinsic magnitude
%from~\cite{bib:rich} at face value (and in the absence of more precise measurements, we have no choice but to do so), 
%we must assume that there could be Type II-P supernovae that are very bright.  This makes it abandonly clear 
%that there exists a very real need to measure the properties of non-Type Ia supernovae with more precision, 
%a task that is ideally suited for existing and planned ground-based supernova surveys.  Second,
%most of the Type Ia's that have low values of  ${\rm log}\mathcal{R}$ (\emph{e.g.}, $<$ 20) 
%tend to be ``early'' candidates, 
%caught near the beginning of their lifetime.  The resulting data have low signal-to-noise ratios, making it
%particularly difficult to discriminate Type Ia's from other types.  
%can make a plot of this using find_bad_Ia.pl & plot_sn.kumac

\begin{figure}[tbh]
\begin{center}
\epsfxsize=3.0in\epsffile{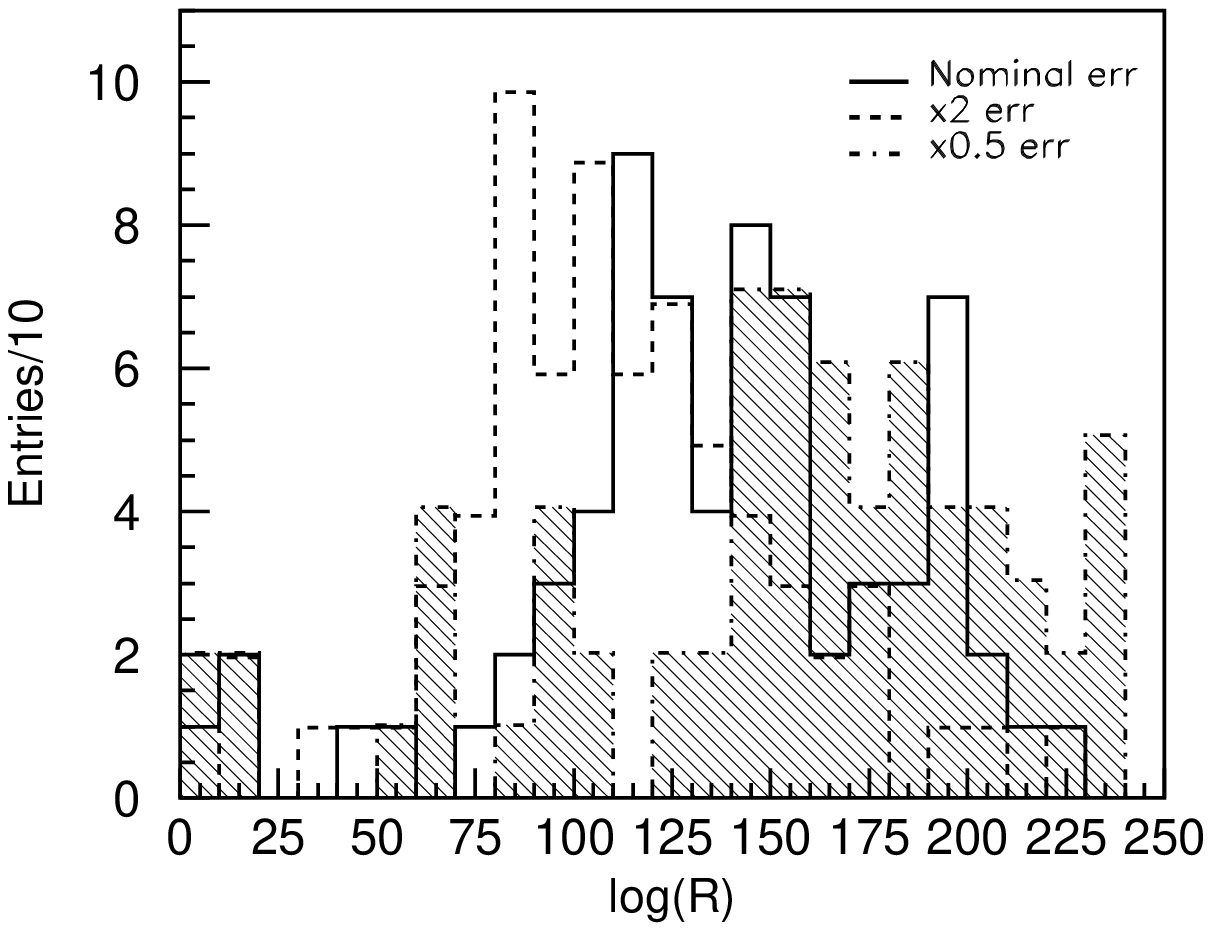}
\end{center}
\caption {
Distributions of ${\rm log}\mathcal{R}$  for simulated Type Ia candidates for nominal flux
errors (solid line), flux errors increased by a factor of 2 (dashed line), and flux errors 
decreased by a factor of 2 (dot-dashed line, filled histogram).  
The histograms have been normalized to the same area to aide in the comparison of their shapes.
}
\label{fig:errs}
\end{figure}

We further check the behavior of $\mathcal{R}$ by 
varying the errors on the fluxes of the simulated SNe Ia to ensure that 
$\mathcal{R}$ changes in the right direction.
This check, always a good idea for a newly introduced statistic,
is particularly important for this Bayes factor, which 
makes use of improper priors that can lead to 
non-intuitive behavior (~\cite{Berger:2001}).
Figure~\ref{fig:errs} shows
the distribution of ${\rm log}\mathcal{R}$ for the ``nominal'' flux errors and for flux errors artificially
increased and decreased by a factor of 2.  Increasing the flux
errors shifts the distribution to the left (\emph{i.e.}, the discrimination power decreases), while
decreasing the flux errors shifts it to the right (\emph{i.e.}, the discrimination power increases).  
This is the expected behavior for a correctly computed  $\mathcal{R}$.  

%%%%%
\subsection{Including Prior Knowledge on Non-Ia Supernova Types}
%%%%%
% see calc_num_withII-P.pl in Supernova_research
So far, we have assumed no prior knowledge of SNe models 
that may contribute to the set of observed non-SNe Ia.  We showed
that the Bayes factor described above is capable of discriminating 
between SNe Ia and non-SNe Ia as well as random unimodal light curves 
that mimic anomalous candidates.
This discrimination, which does not require either the knowledge of a complete 
set of objects that can mimic an SN Ia signal or the knowledge of the possible
behavior of anomalous non-supernova objects that can contaminate an SN Ia
signal, is good.  
As Fig.\,\ref{fig:IaandII-P} shows, we could simply define
a polynomial cut on $\mathcal{R}$ as a function of $\mathcal{R}$
and have a very good discriminant.

However, one might be interested in considering a Bayes factor for which
all \emph{known} non-SN Ia candidates would have $\mathcal{R}<1$.
First, it is simply better to have a $\mathcal{R}$ that behaves intuitively.
Second, it is obvious that including more prior knowledge (\emph{i.e.}, the
knowledge of what can potentially mimic an SN Ia signal), can only 
sharpen the discrimination between SNe Ia and non-SNe Ia.
Third, and more importantly, it is a good idea to 
have a discriminant such that $\mathcal{R}$ $<$ 1 when the candidate is more 
likely to be an SN Ia than not, and $\mathcal{R}$ $>$ 1 otherwise. 
This allows one to use a formalism similar to that described in~\cite{Wald:1945,Wald:1947} in order
to set thresholds on $\mathcal{R}$ with meaningful,
pre-determined Type I and II error rates.

Explicitly including the knowledge about the behavior of non-SNe Ia into 
Eqn.~\ref{eqn:general}, we define:
\begin{tiny}
\begin{eqnarray}
\mathcal{R^{\prime}}= &\nonumber \\
\frac{P(\text{Phot}|\text{Ia})}{P(\text{Phot}|\text{II-P})P(\text{II-P}|\text{non-Ia}) + P(\text{Phot}|\text{Ibc})P(\text{Ibc}|\text{non-Ia}) + P(\text{Phot}|\text{IIn})P(\text{IIn}|\text{non-Ia}) + P(\text{Phot}|\text{anything})P(\text{anything}|\text{non-Ia})},
\label{eqn:general2}
\end{eqnarray}
\end{tiny}
where the denominator now accounts for 
the probability that the observed photometry can come from a Type II-P supernova, 
$P(\text{Phot}|\text{II-P})$; a Type Ibc supernova, $P(\text{Phot}|\text{Ibc})$;
or from a Type IIn supernova, $P(\text{Phot}|\text{IIn})$.
$P(\text{Phot}|\text{anything})$ is equivalent to the denominator in Eqn.\,\ref{eqn:R_final2}.
Note that the prior terms in Eqn.~\ref{eqn:general2} are such that 
\begin{eqnarray}
P(\text{II-P}|\text{non-Ia})=P(\text{Ibc}|\text{non-Ia})=P(\text{IIn}|\text{non-Ia})=P(\text{anything}|\text{non-Ia})=\frac{1}{4}
\end{eqnarray}
In other words, there is an equal probability of
measuring a Type Ia supernova, a Type Ibc supernova, a Type II-P supernova
or some other object denoted as ``anything''.
It is of course trivial to introduce relative rates if they are known;
however, it is immaterial for our purpose, which is demonstrating 
the performance of the method.

\begin{figure}[!htb]
 \begin{center}
 $\begin{array}{c@{\hspace{0.0in}}c}
\epsfxsize=3.0in\epsffile{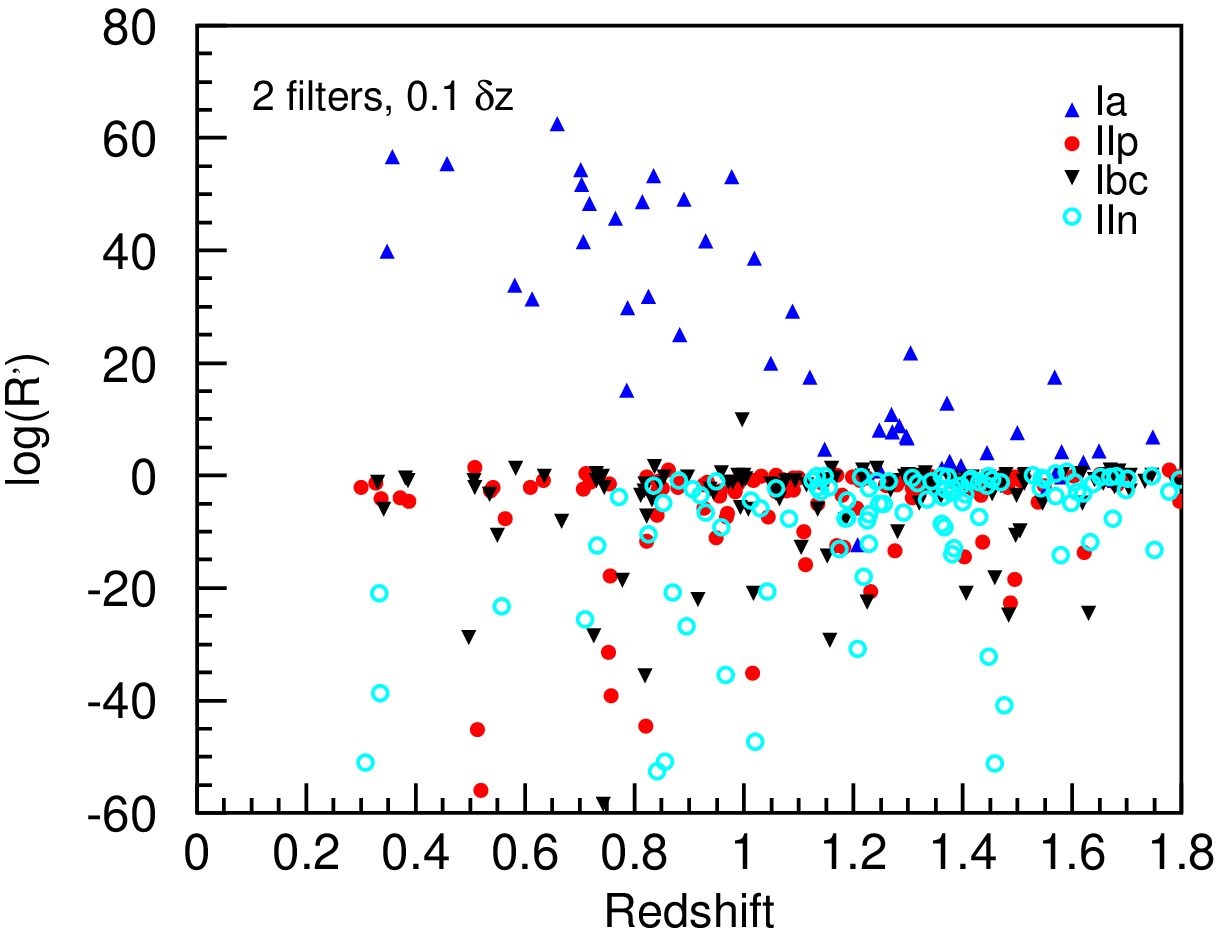} & 
\epsfxsize=3.0in\epsffile{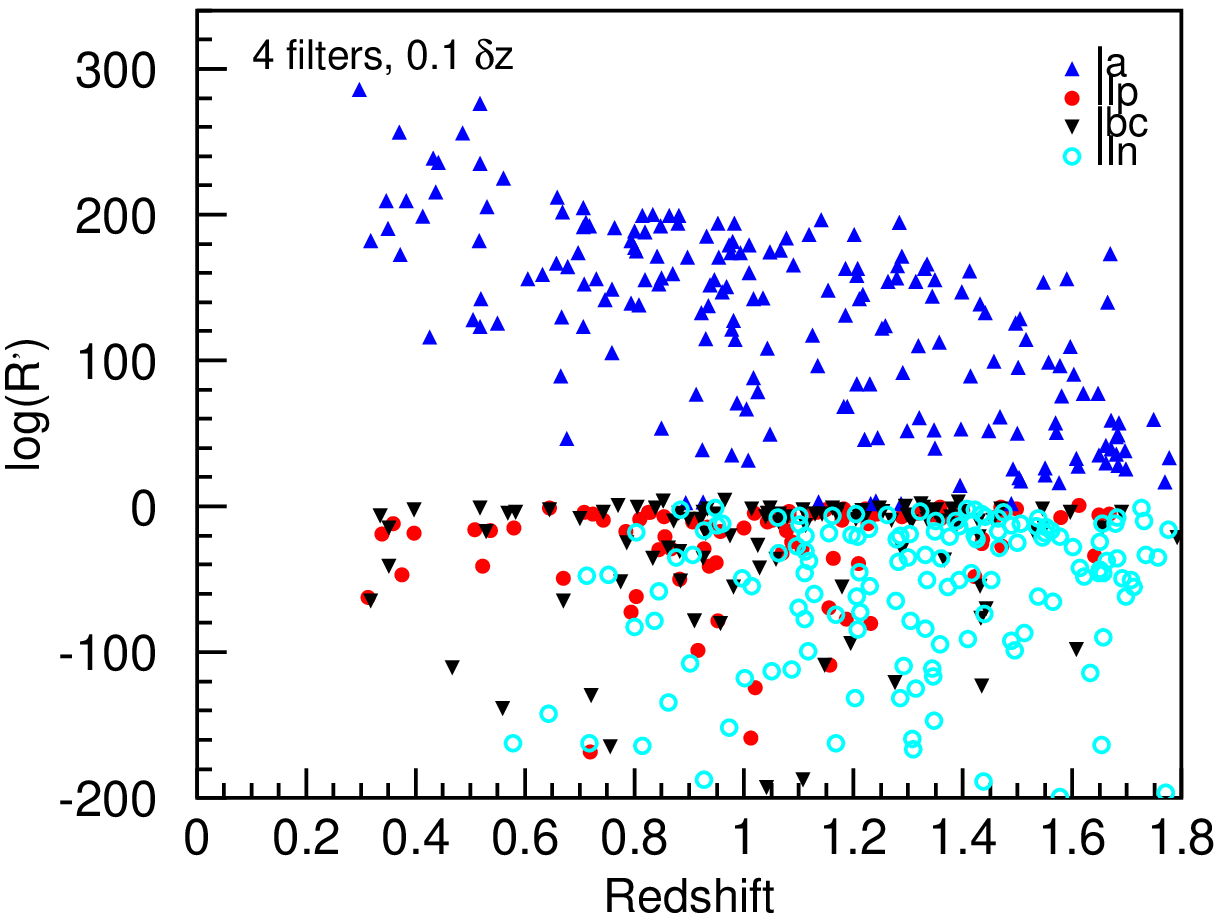} \\
\epsfxsize=3.0in\epsffile{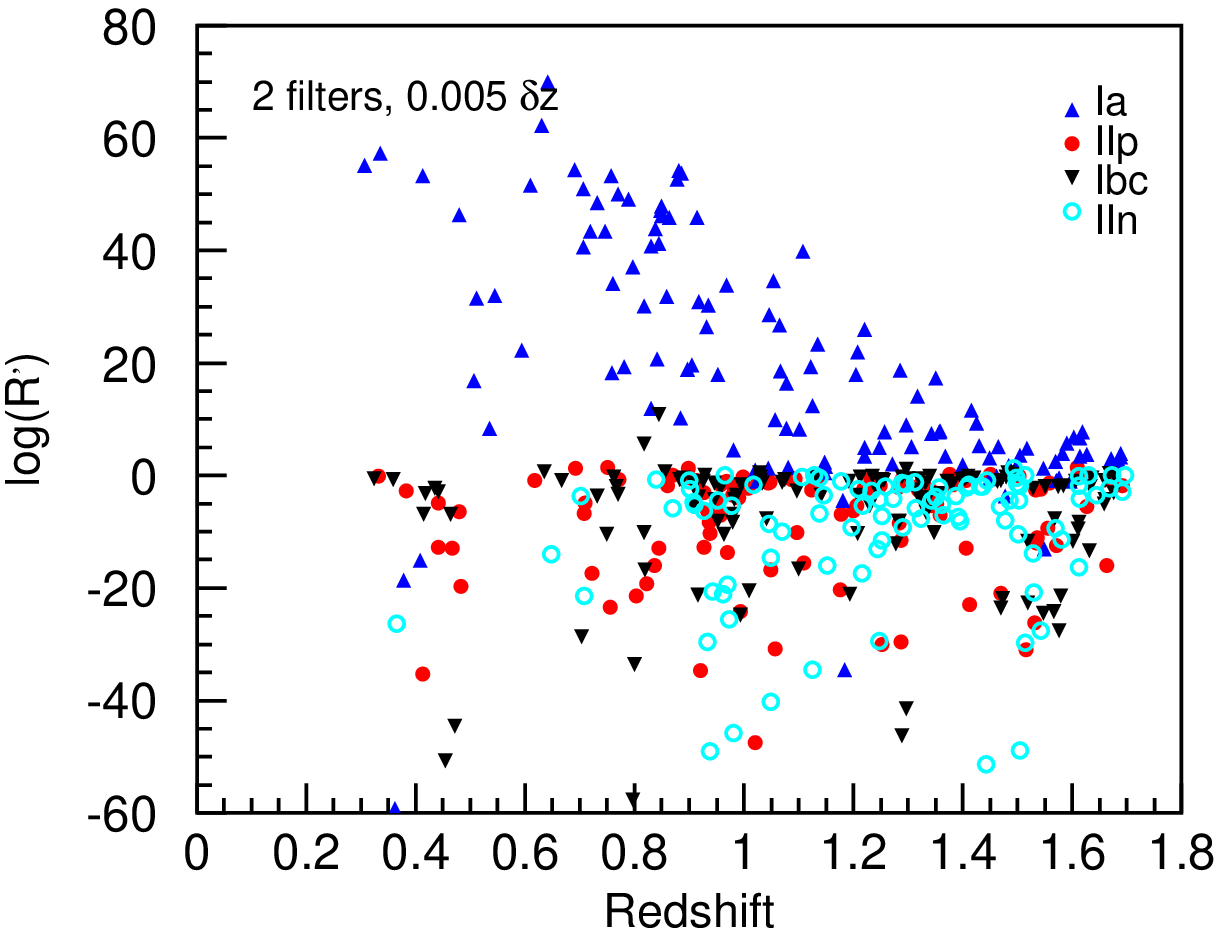} & 
\epsfxsize=3.0in\epsffile{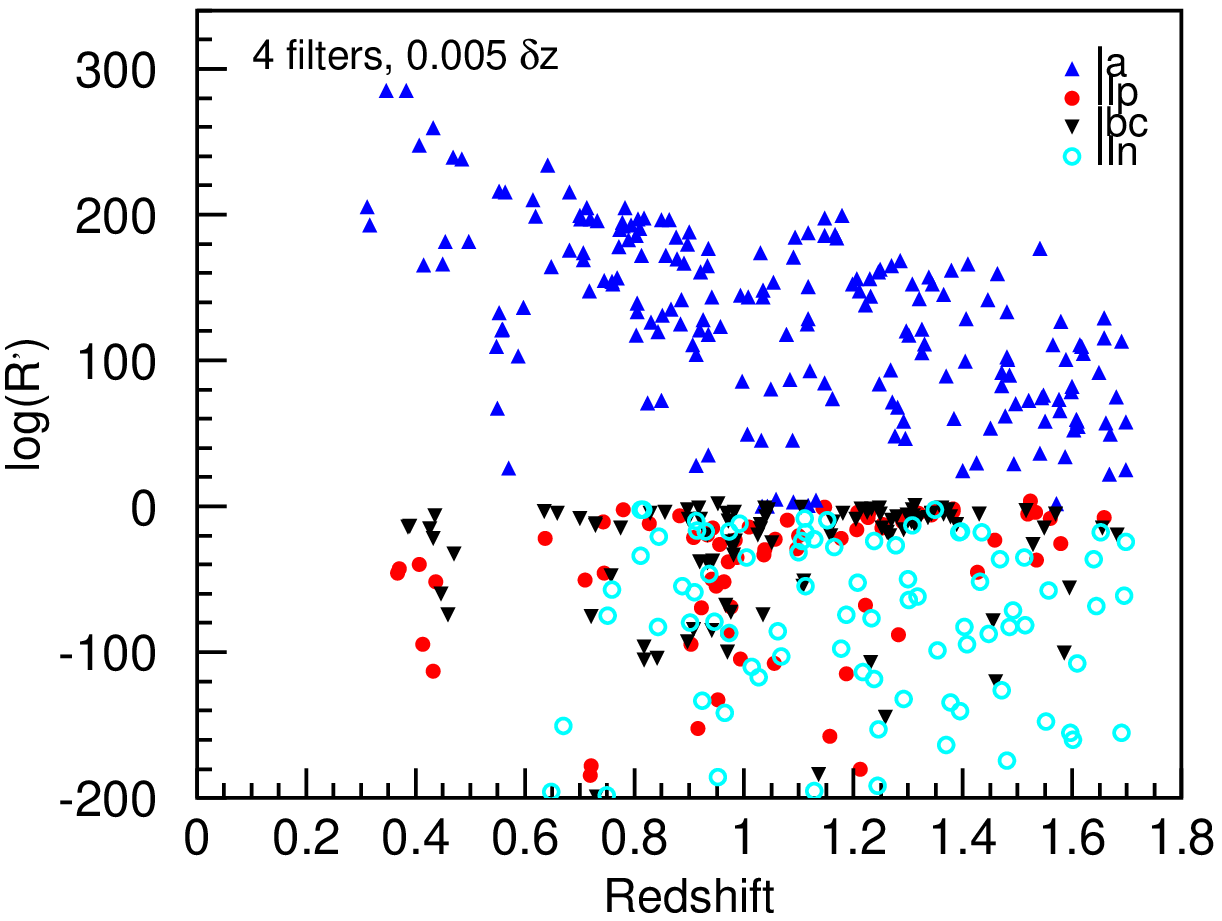} \\
\end{array}$
  \end{center}
\caption[]
{\label{fig:IaandII-Ppriorknowledge}
Distributions of ${\rm log}\mathcal{R^{\prime}}$ vs. redshift for Type Ia's (upward turned triangles),
Type Ibc's (downward turned triangles),
Type II-P's (filled circles) and Type IIn's (open circles), with a 0.1 error
on the candidate redshifts (top row) and a 0.005 error on the candidate
redshifts (bottom row), for 2 filters (left plots) and 4 filters (right plots).
}
\end{figure}

The non-Type Ia probabilities are calculated in exactly the same way as those for Type Ia's, 
using the available models for the corresponding types.
The distribution of  ${\rm log}\mathcal{R^{\prime}}$  vs. redshift
in shown in Fig.~\ref{fig:IaandII-Ppriorknowledge}, for the case of  
2 filters (left column) and 4 (right column) filters with 0.005 flux errors (bottom row), 
and 0.1 errors (top row) on the redshift.
Randomly generated, uni-modal data all have large negative values $log(\mathcal{R})$'s 
that dwarf the scales on these figures.
The errors on the flux are those expected for a JDEM mission described above.
Compared Fig.~\ref{fig:IaandII-Ppriorknowledge} 
to Fig.\,\ref{fig:IaandII-P}, it is clear that not only has the discrimination 
between SNe Ia and everything else increased, but also the 
SNe Ia generally have $log(\mathcal{R}^\prime)>0$
and other candidates have $log(\mathcal{R}^\prime)<0$.  This is the desired behavior for
${\mathcal{R}^\prime}$.
%%%

A number of features of Fig.~\ref{fig:IaandII-Ppriorknowledge} are similar to those
apparent in Fig.~\ref{fig:IaandII-P}, such as the increase in the discrimination power
when more and/or better information becomes available.  

%It is thus apparent that including some prior information about the possible sources of background into 
%the calculation of $\mathcal{R}$ does improve the discrimination.  This will be further explored
%in a future paper~\citep{bib:min}.

%%%%%
\subsection{Dangers of a Finite Set Assumption}
\label{sec:danger}
%%%%%
As we explained in Section~\ref{section:intro}, existing Bayesian-based methods
of supernova classification assume a finite set of possible objects 
that can mimic an SN Ia signal.  To demonstrate the danger of this limiting
assumption, we use our unimodal fake light curves and calculate Bayes factors
defined as:
\begin{small}
\begin{eqnarray}
R_{Ia} = \frac{P(\text{Phot}|{\rm Ia})P({\rm Ia})}{P(\text{Phot}{\rm |Ibc})P({\rm Ibc}|{\rm non-Ia})+P(\text{Phot}{\rm |II-P})P({\rm II-P}|{\rm non-Ia})+P(\text{Phot}{\rm |IIn})P({\rm IIn}|{\rm non-Ia})}
\end{eqnarray}
\end{small}
and 
\begin{small}
\begin{eqnarray}
R_{II-P} = \frac{P(\text{Phot}|{\rm II-P})P({\rm II-P})}{P(\text{Phot}{\rm |Ibc})P({\rm Ibc}|{\rm non-Ia})+P(\text{Phot}{\rm |Ia})P({\rm Ia}|{\rm non-Ia})+P(\text{Phot}{\rm |IIn})P({\rm IIn}|{\rm non-Ia})}
\end{eqnarray}
\end{small}

\begin{figure}[tbh]
\begin{center}
\epsfxsize=4.0in\epsffile{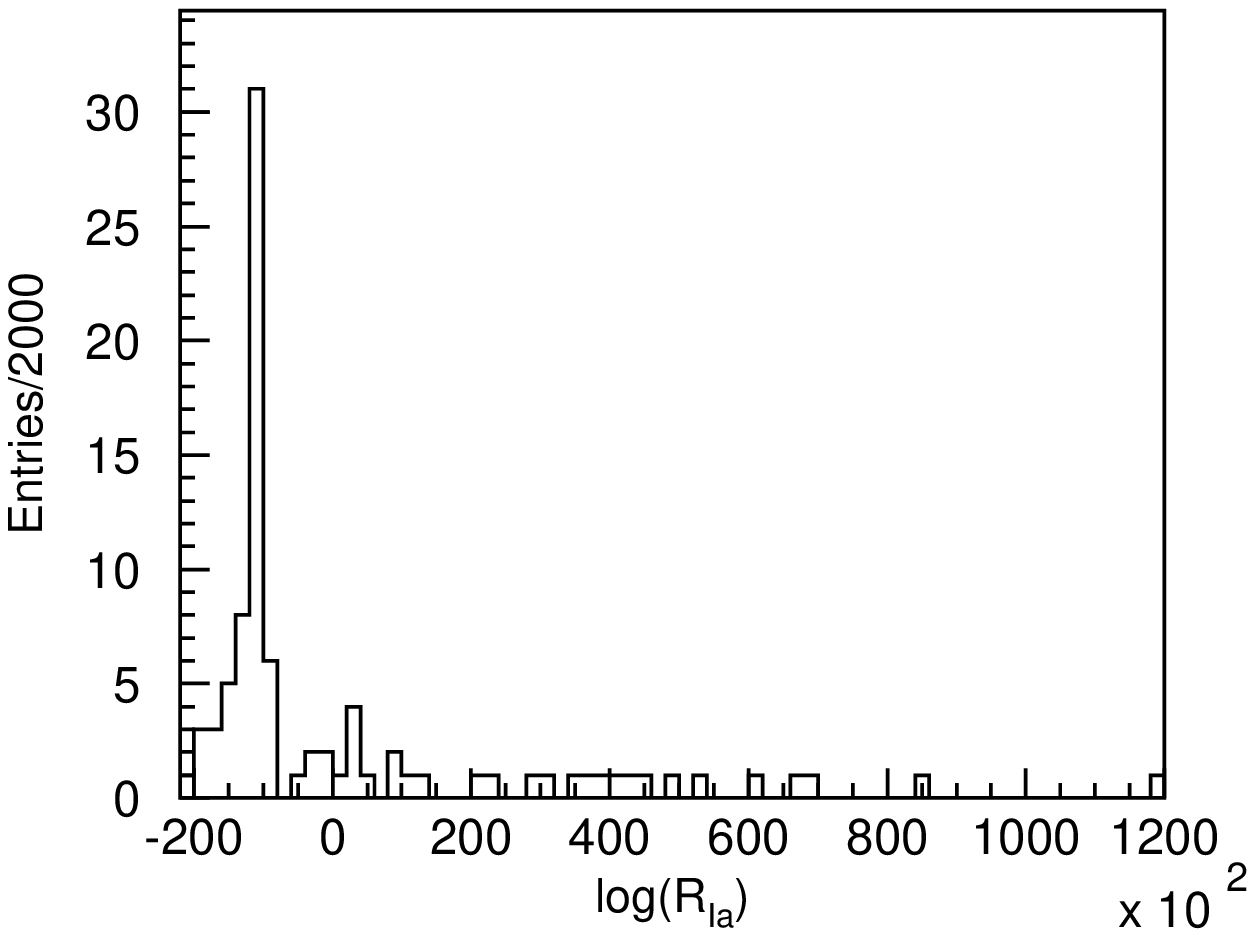} \\
\epsfxsize=4.0in\epsffile{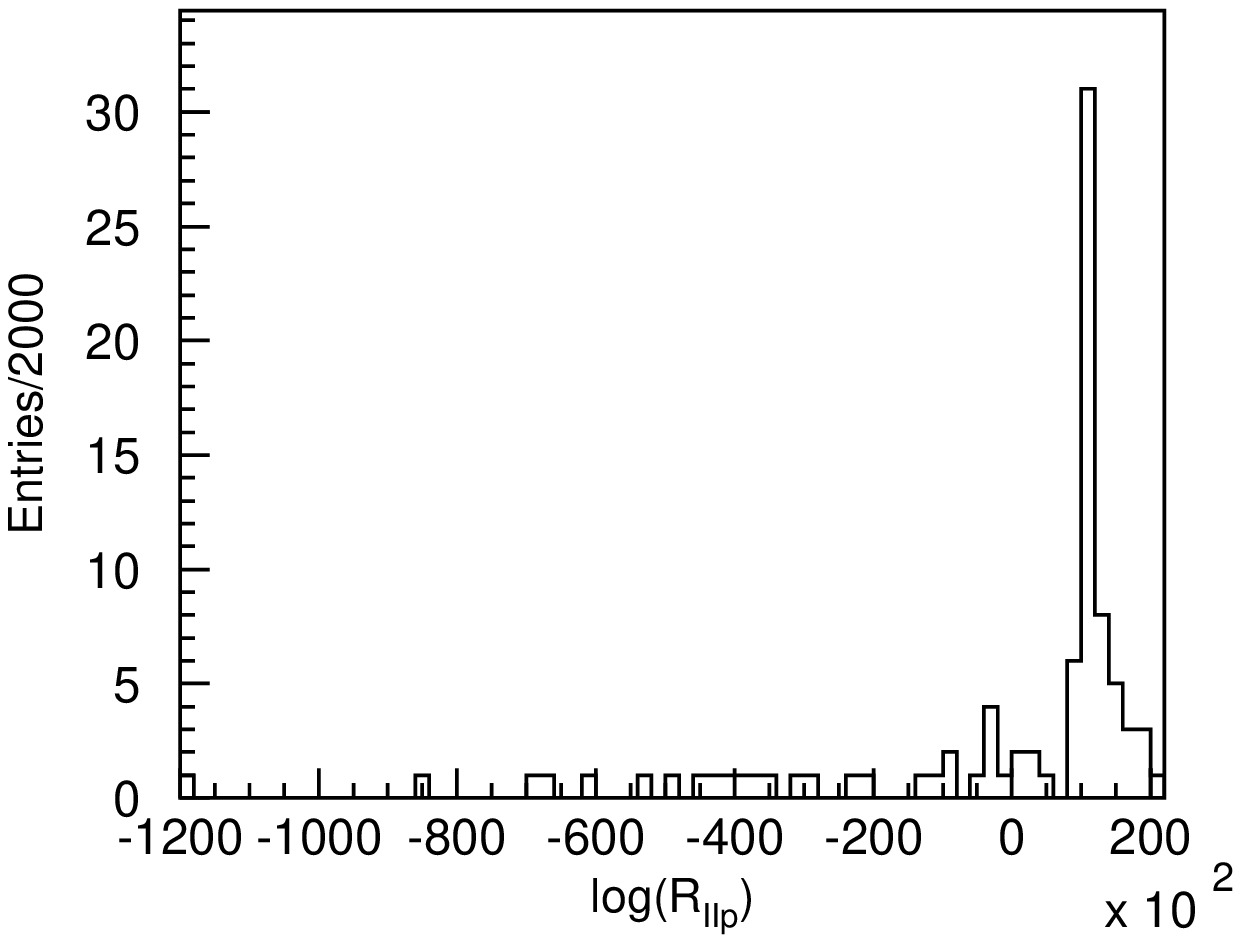} 
\end{center}
\caption {\label{fig:naive}  
The distributions of ${\rm log}R_{Ia}$ (top) and ${\rm log}R_{II-P}$ (bottom)
for a set of unimodal random light curves.
}
\end{figure}

Figure\,\ref{fig:naive} shows that the fake objects can be mis-identified 
as supernovae of types other than ia (in particular, as Type II-P's), as
well as SNe Ia: there are candidates with $R_{Ia}$ $>$ 1.  
This further demonstrates the need for a general a
formalism that is  capable of discriminating 
a certain type of 
supernovae (most often Type Ia's) from \emph{anything else}.

%%%%%%%%%%%%%%%%%%%%%%%%%
\section{Summary}
\label{section:disc}
%%%%%%%%%%%%%%%%%%%%%%%%%
We have introduced a new photometric supernova classification scheme that uses a Bayes factor
based on color.  The proposed method is fundamentally different 
from previous supernova classification methods including our own~\citep{bib:kuzn} 
because it allows one to discriminate not only between supernovae of different
types but also  between supernovae and ``anomalous'' objects.  It
has a number of definite advantages over many existing techniques
for selecting SNe Ia out of a pool of supernova candidates.  The main one is that 
it does not pre-suppose any prior knowledge about the objects that could potentially 
mimic a Type Ia signal.  It can thus be used as a very good first-pass Type Ia 
classifier.  With the current poor knowledge of the behavior of non-Type Ia supernovae, 
especially at high redshifts, and the expected dramatic increase in the discoveries
of new, as yet unknown classes of transient astronomical objects, this feature of
the method will be invaluable for future supernova surveys.   This is not, however,
an excuse not to obtain as much information about non-Type Ia supernovae as one 
possibly can, as evidenced by the advantage of computing $\mathcal{R}^\prime$,
which includes information about the light curves of Type Ia, Ibc, IIn, and II-P supernovae
(obviously, more information means a better performance).

Another principal advantage of the proposed method is that 
if the Bayes factor described in Section~\ref{section:results}
is used as a discriminant, the only supernova models that are required
are those for SNe Ia, which are the best studied and most complete of all the supernova types.
One may argue that the same is true for a $\chi^2$-based method.  
However,  $\chi^2$ methods suffer from a number of problems 
described in Section\,\ref{section:intro},
the least of which is that for the case of data with large uncertainties
a given supernova candidate will appear to agree with many possible supernova type hypotheses,
with one necessarily giving the ``best'' $\chi^2$ (however insignificant 
the difference between this best $\chi^2$ and the $\chi^2$'s from the other fits 
may be).  The Bayes factor, on the other hand, would be of order 1, 
reflecting ambiguity between the hypothesis that the candidate is an SN Ia
and that it is anything else.

Finally, as it is a Bayesian approach, it accounts for all
systematic and statistical uncertainties of the measurements 
in all their forms.  

%For Type Ia supernovae photometric data in particular, 
%a number of conclusions can be drawn from these studies:
We demonstrated the method 
on a simulated dataset expected from a possible future large-scale space-based
JDEM.  It must be noted that most existing data sets include primarily SNe Ia;
at the present time
large datasets with exotic supernovae that would allow
us to perform a statistically meaningful study of the method's performance are not available
(although we are planning on exploring the application of this method to an
existing dataset of non-standard
supernovae in a future publication).
Nevertheless, a number of interesting conclusions can be drawn from the simulation studies
described above:
\begin{itemize}
\item
The method provides a 100\% discrimination between SNe Ia and unimodal random data.  
This is encouraging, since many transient objects (such as active galactic nuclei) that are
sometimes mistaken for supernovae tend to have a rather erratic behavior, deviating
far more from an SN Ia light curve than the simple random unimodal data.
More fundamentally, the discrimination also shows that $\mathcal{R}$ 
and $\mathcal{R}^\prime$ behave as expected.

\item
The discrimination between Type Ia's and other supernova types is near 100\% at low redshifts
when calculating log$\mathcal{R}$.  
At higher redshifts, Fig.~\ref{fig:IaandII-P} shows that 
that a ``straight line'' 
cut on ${\rm log}\mathcal{R}$ would render a reasonably high 
purity and efficiency for SNe Ia.
Alternatively, one could invent a more sophisticated
cut (\emph{e.g.}, a polynomial).
The discrimination is practically 100\% at all redshifts if
log$\mathcal{R}^\prime$ is used (\emph{i.e.}, when information about 
the behavior of Type II-P, Type Ibc, and Type IIn supernovae
is included in calculating $P(\text{Phot}|\text{non-Ia})$).

\item  The method performs better when more information about the data
becomes available.  For example, 
the discrimination between SNe Ia and non-SNe Ia increases 
dramatically when the number of filters goes from 2 to 4.
This shows that, despite the use of improper priors, the 
Bayes factor behaves properly.

\item  Increasing the precision on the redshift of the supernova candidates 
also improves the discrimination between Type Ia's and non-SNe Ia.

\item It is important that the data and the models used have a good
match in terms of expected colors.  For example, if the extinction
assumptions in the data are different from those in the data, 
real SNe Ia are more likely to be classified as anomalous objects.

\item Figures\,\ref{fig:IaandII-P} 
and \ref{fig:IaandII-Ppriorknowledge} indicate that 
there is sufficient separation between various supernova candidates, so that this method 
could be used for classification in the strict sense of giving a type (\emph{e.g.}, a Type Ia, 
a Type Ibc, \emph{etc.})
to each candidate, along with some associated probability.
\end{itemize}

The method does not make use of magnitudes.  One
might see this as an advantage if one does not
entirely trust the distribution of intrinsic magnitudes.
Insertion of absolute magnitudes is possible, but 
the formulation of $\mathcal{R}$ ($\mathcal{R}^\prime$) becomes inelegant 
and requires knowledge of an upper limit of intrinsic 
magnitudes of anomalous candidates that, by definition,
have not been observed.  It is indeed fortunate that color alone is
sufficient to classify supernovae, allowing for a simple and elegant
solution for  $\mathcal{R}$ ($\mathcal{R}^\prime$).

There is another reason not to include magnitudes into the formalism,
at least at the present time.  The distributions of the intrinsic magnitudes
for non-SNe Ia are at the moment rather poorly known~\citep{bib:rich}.
In fact, very little is known about high-redshift non-SNe Ia (for example,~\cite{dahlen}
remains the only measurement of the non-SNe Ia rates to redshifts of $\sim$ 1).
There exists a very real need to measure the properties of non-SNe Ia supernovae with more precision, 
a task that is ideally suited for existing and planned large-scale supernova surveys.  

%Second,
%most of the Type Ia's that have low values of  ${\rm log}\mathcal{R}$ (\emph{e.g.}, $<$ 20) 
%tend to be ``early'' candidates, 
%caught near the beginning of their lifetime.  The resulting data have low signal-to-noise ratios, making it
%particularly difficult to discriminate Type Ia's from other types.  
%can make a plot of this using find_bad_Ia.pl & plot_sn.kumac

It is also important to note the computational challenges
in calculating $\mathcal{R}$ ($\mathcal{R}^\prime$).
The number of filters used in the calculation depends on the precision needed for the 
integration of $\{f^{\prime\,k}_j\}$.  
It was found that for 4 filters about 150 integration points for $f^{\prime\,k}_j$
were needed for a precise calculation of the 
denominator, $P(\text{Phot}|\text{non-Ia})$.  This was found
by increasing the number of integration points
until the value of the denominator became stable.

In order to complete the computations in a reasonable amount of time,
it was necessary to perform the calculations of the Bayes factor for many candidates in parallel.
The computational feasibility also depended on approximating
$b\rightarrow b_i$, as was discussed in Section~\ref{section:details}, so as to reduce
the number of integrations from $N \times M$ (where $N$ is the number of flux measurements and
$M$ is the number of filters) to $M$.
This effectively required that we gave up 
information about the supernova colors
measured in the various filters and allowed the colors
to vary measurement-to-measurement.  One might be concerned that 
this approximation would in fact allow
for a greater diversity in what is considered an SN Ia -- that is, in general objects would
have a greater chance of faking an SN Ia.  However, in our studies we found 
that, at least with the level of precision of the models and simulated data used, 
the calculated Bayes factor provided desired discrimination between SNe Ia
and other objects, leading us to believe that this is not a significant effect.

We also point out that our proposed technique is 
general enough to be used for objects other than supernovae.
For example, ~\cite{richards} 
propose a Bayesian classifier to differentiate
quasars and stars, defined as:
\begin{eqnarray}
P(\text{star}|x)=\frac{P(x|\text{star})P(\text{star})}
{P(x|\text{star})P(\text{star})+P(x|\text{quasar})P(\text{quasar})}
\end{eqnarray}
where $x$ is a candidate's position in a 4-dimensional 
color space.  Although the likelihoods, $P(x|\text{star})$
and $P(x|\text{quasar})$ are obtained using a training 
sample derived from real data, new data inevitably bring
new objects which could be accounted for by inserting 
an ``anomaly'' term
similar to $P(\text{Phot}|\text{anything})P(\text{anything})$ 
in Eqn.\,\ref{eqn:general2}.
This term would sweep up those objects that do not conform
to the existing models for quasars or stars in the same way that 
$P(\text{Phot}|\text{anything})P(\text{anything})$
accounts for anomalous supernova candidates;
its exact form would depend on the nature of the statistical fluctuations
in the data.

%%%%%%%%%%%%%%%%%
\subsection{$\mathcal{R}$ Used in the Context of an SN Ia Trigger}
\label{section:trigger}
%%%%%%%%%%%%%%%%%%%
Note that while our method relies only on photometric information about supernovae,
some proposed future space-based dark energy missions do plan on 
obtaining the spectrum of every candidate.  One possible scenario would be
to obtain the spectrum of a candidate  
provided that a) it is highly likely to be an SN Ia, and b) it is at its peak brightness.
In this case it is necessary to have reliable means to be able to tell
whether or not a given candidate is most likely an SN Ia or not based
on its pre-maximum photometric measurements alone.  In other words, it is important to have
a \emph{trigger} mechanism that would photometrically select candidates for possible 
spectroscopic follow-up.  Our proposed Bayes factor can be simply modified
to allow for such a usage.  
Assuming that one wants to trigger on supernovae 
before maximum brightness, the Bayes factor becomes:
\begin{tiny}
\begin{eqnarray}
\mathcal{R^{\prime\prime}}=
\frac{P(\text{Phot}|\text{Ia pre-max})}{P(\text{Phot}|\text{Ia post-max})+P(P(\text{Phot}|\text{II-P})P(\text{II-P}|\text{non-Ia}) + P(\text{Phot}|\text{Ibc})P(\text{Ibc}|\text{non-Ia}) + P(\text{Phot}|\text{anything})P(\text{anything}|\text{non-Ia})},
\label{eqn:general3}
\end{eqnarray}
\end{tiny}
where $P(\text{Phot}|\text{Ia pre-max})$ would only include supernova
models with points before maximum light, while 
$P(\text{Phot}|\text{Ia post-max})$
would  only include those with points after maximum light.
After calculating this Bayes factor,
a cut would be made at, say, $R^{\prime\prime}>1$
to choose those candidates that are likely to be SNe Ia and that have not
yet reached maximum.  

Better still, one could use a \emph{sequential analysis}
technique~\citep{Wald:1945,Wald:1947} 
to minimize the data required to make this decision
while simultaneously controlling  identification errors.
This is done by setting thresholds on $\mathcal{R^{\prime\prime}}$
based on pre-selected Type I and Type II error rates.
The demonstration of the performance of a sequential analysis-based 
approach will be the subject of a future publication.

%%%%%%%%%%%%%%%%%%%%%%%%%%%
\acknowledgements
\section{Acknowledgments}
%%%%%%%%%%%%%%%%%%%%%%%%%%
NC gratefully acknowledges support from a Cottrell College Science Award from the 
Research Corporation and from NSF RUI award \#0806877, as well as from Hamilton College.  
BC is partially supported by grant \#DE-FG02-95ER40893 from the U.S. Department of Energy.
We thank Steve Young
at Hamilton College for computer technical support.

\end{document}